\documentclass[12pt, draftclsnofoot, onecolumn]{IEEEtran}
\usepackage{cite}
\usepackage{graphicx}
\usepackage{amsmath}
\usepackage{array}
\usepackage{mdwmath}
\usepackage{amssymb}
\usepackage{mdwtab}
\usepackage{stfloats}
\usepackage[tight,footnotesize]{subfigure}
\usepackage{amsmath,amsthm}
\usepackage{threeparttable}
\usepackage{color}
\usepackage{url}
\usepackage{algpseudocode}
\usepackage{algorithm}
\usepackage{multicol}
\usepackage{bm} 
\usepackage{mathabx}
\usepackage{tabularx}
\algrenewcommand{\algorithmicrequire}{\textbf{Input:}}
\algrenewcommand{\algorithmicensure}{\textbf{Output:}}

\newtheorem{example}{Example}
\newtheorem{theorem}{Theorem}
\newtheorem{remark}{Remark}
\newtheorem{lemma}{Lemma}

\begin{document}

\title{\LARGE Spatial Scattering Modulation with Multipath Component Aggregation Based on Antenna Arrays}
\author{\small
Jiliang~Zhang,~\textit{Senior Member~IEEE},~Wei~Liu,~\textit{Senior Member IEEE},
~Alan~Tennant,~\textit{Senior Member IEEE},
Weijie Qi, Jiming Chen, and Jie~Zhang,~\textit{Senior Member IEEE}
\thanks{
Jiliang Zhang is with College of Information Science and Engineering, Northeastern University, Shenyang, China.
Jie Zhang is with the Department of Electronic and Electrical Engineering, The University of Sheffield, Sheffield, UK, and also with Ranplan Wireless Network Design Ltd., Cambridge, UK.
Wei Liu and Alan Tennant are with the Department of Electronic and Electrical Engineering, The University of Sheffield, Sheffield, UK. 
Weijie Qi and Jiming Chen are with Ranplan Wireless Network Design Ltd., Cambridge, UK.
}
\thanks{This research was a part of KTP project - ``Massive MIMO Beamforming for 5G'', which is funded by UKRI through Innovate UK.}
}
\markboth{IEEE}%
{Shell \MakeLowercase{\textit{et al.}}: Bare Demo of IEEEtran.cls for Journals}
\maketitle
\vspace{-0.5in}
\begin{abstract}
In this paper, a multipath component aggregation (MCA) mechanism is introduced for spatial scattering modulation (SSM) to overcome the limitation in conventional SSM that the transmit antenna array steers the beam to a single multipath (MP) component at each instance. In the proposed MCA-SSM system, information bits are divided into two streams. One is mapped to an amplitude-phase-modulation (APM) constellation symbol, and the other is mapped to a beam vector symbol which steers multiple beams to selected strongest MP components via an MCA matrix. In comparison with the conventional SSM system, the proposed MCA-SSM enhances the bit error performance by avoiding both low receiving power due to steering the beam to a single weak MP component and inter-MP interference due to MP components with close values of angle of arrival (AoA) or angle of departure (AoD). For the proposed MCA-SSM, a union upper bound (UUB) on the average bit error probability (ABEP) with any MCA matrix is analytically derived and validated via Monte Carlo simulations. Based on the UUB, the MCA matrix is analytically optimized to minimize the ABEP of the MCA-SSM. Finally, numerical experiments are carried out, which show that the proposed MCA-SSM system remarkably outperforms the state-of-the-art SSM system in terms of ABEP under a typical indoor environment.

\end{abstract}


\IEEEpeerreviewmaketitle

\section{Introduction}
Index modulation (IM) with sparse symbol mapping has been considered as a promising candidate to facilitate 5G/B5G point-to-point connections \cite{SpatialModulation, IM1, IM2, IM3, IM4, polarization-space, polarsk, SSG}.
Recently, spatial scattering modulation (SSM) has been proposed as an attractive solution for  IM with massive multiple-input-multiple-output (MIMO) antenna arrays at both the transmitter (Tx) and the receiver (Rx) \cite{Spatial,BIM,Millimeter}. The SSM system requires fewer radio frequency (RF) chains at the Tx to reduce power consumption.

In a conventional SSM system, the information bits are separated into two streams. 
The first stream is mapped to an amplitude-phase-modulation (APM) constellation symbol; 
the second stream is mapped to a single multipath (MP) component out of several MP components with the largest gains in the wireless channel according to the input bits. Recently, SSM has drawn much attention as a strong candidate for 5G/B5G modulation systems
with various extensions such as the generalized SSM \cite{Generalized, PY5}, 3-D GSSM  \cite{YJiang}, polarized SSM \cite{Polarized}, and adaptive SSM \cite{Adaptive}. The diversity order of the conventional SSM system was analytically derived in  \cite{Diversity}.

However, the conventional SSM system suffers from two problems. 
One is the risk of low receiving power as the transmitter may steer the beam to an MP component with a low cluster gain; 
the other is the inter-MP interference because the values of angle of arrival (AoA)/angle of departure (AoD) for different MP components could be too close for the receiver/transmitter to resolve. 
In \cite{PY3, PY4}, signal shaping mechanisms are introduced to address the above problems via a recursive optimization approach
at the cost of a heavy computational burden.

Therefore, the main objective of this paper is to propose an analytic MP component aggregation (MCA) mechanism in the design of SSM to overcome both above problems. 
On the one hand, the proposed MCA-SSM increases the possible minimal received power by aggregating multiple MP components 
for an overall large gain. On the other hand, MP components are aggregated in an orthogonal manner to avoid selecting two MP components with close AoAs or AoDs. Moreover,  an analytic MCA matrix derivation approach is presented instead of
through
 recursive optimization, which leads to reduced computational complexity for its implementation.
Therefore, the proposed MCA-SSM is expected to outperform the conventional SSM system in terms of average bit error probability (ABEP). 

\begin{figure}[h]
\centering
\includegraphics[scale=0.8]{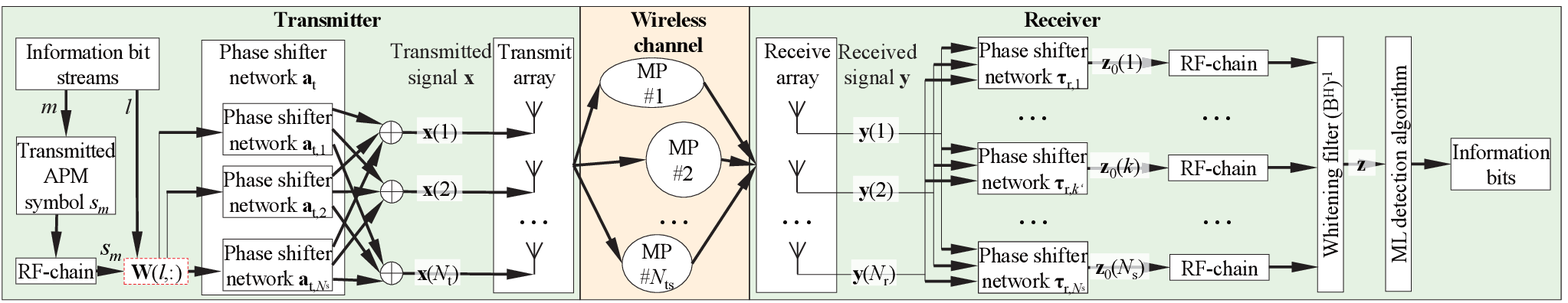}
\caption{System model of the proposed MCA-SSM.} \label{fig1}
\includegraphics[scale=0.8]{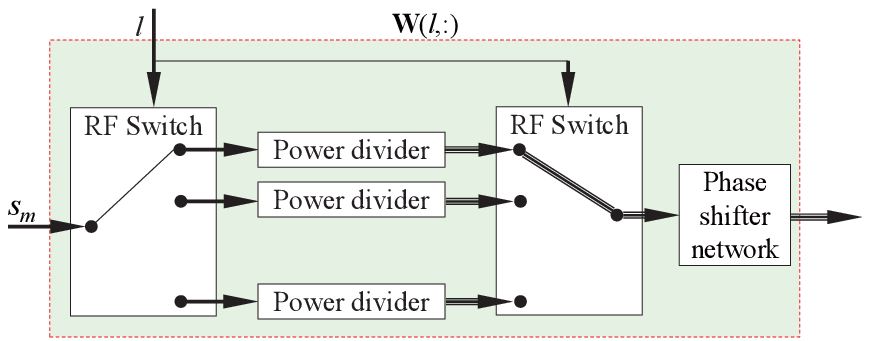}
\caption{The MCA module, the red rectangular in Fig. \ref{fig1}.} \label{fig2}
\vspace{-5mm}
\end{figure}

The main contributions of this paper are summarized as follows:
\begin{itemize}
\item  An MCA mechanism is introduced in the state-of-the-art SSM system, as shown in Fig. \ref{fig1}. Therein, MP components in the wireless channel can be aggregated via the MCA matrix $\mathbf{W}$ into several beam patterns, as shown in Fig. \ref{fig2}. At each instance, one of the beam patterns, associated with one row of the MCA matrix, is selected according to the input bits. 
\item To facilitate optimization of the bit error performance for the MCA-SSM system, a quick evaluation approach for the MCA-SSM system is designed. More specifically, a tight union upper bound (UUB) on the ABEP for an MCA-SSM system with an arbitrary MCA matrix
is derived and validated via Monte-Carlo simulations. 
\item The optimization problem of the MCA matrix is formulated to minimize the ABEP of the proposed MCA-SSM based on the UUB.
Therein, the MCA matrix is constructed 
pursuing orthogonality to circumvent analytical intractability.
The optimal MCA matrix is analytically derived as a solution to a set of linear equations.
The simple expression of the MCA matrix facilitates easy
implementation of
 MCA-SSM. 
\item A systematic evaluation of the proposed MCA-SSM system is demonstrated in a typical indoor environment based on an
intelligent ray launching algorithm (IRLA).
 Numerical results show that the proposed MCA-SSM system significantly outperforms the conventional SSM system in terms of ABEP.
\end{itemize}

The rest of this paper is structured as follows. 
The system model of the proposed MCA-SSM is described in Section II, where the MCA matrix is defined.
With an arbitrary MCA matrix, the UUB on the ABEP of the MCA-SSM system is derived in Section III.  In Section IV, an analytical MCA mechanism is proposed to solve the optimization problem.
Section V provides a case study on how to apply results in Section IV with given channel parameters. 
A systematic evaluation of the proposed MCA-SSM system in a typical indoor environment is presented in Section VI.
Finally, conclusions are drawn in Section VII.

Notations of this paper are summarized in TABLE I.

\begin{table}[h]
\centering
\caption{Notations in this paper.}
\label{table_param}
\begin{tabular}{l l||l l}
\hline 
\bf{Notations} & \bf{Definitions} &\bf{Notations} & \bf{Definitions}\\ 
\hline
$N_{\mathrm{t}}$ & Number of Tx antenna elements &
$N_{\mathrm{r}}$ & Number of Rx antenna elements\\
$n_{\mathrm{t}}$ & Index of transmit antenna elements&
$n_{\mathrm{r}}$ & Index of receive antenna elements\\
$\mathbf{W}$ & The MCA matrix &
$\mathbf{H}$ & The MIMO channel matrix\\
$N_{\mathrm{ts}}$ & Total number of multipath components&
${\bf{\tilde{H}}}_{n_{\mathrm{s}}}$ & Channel matrix of the $n_{\mathrm s}$-th MP component\\
${\mathbf{a}}_{{\mathrm{t}}}$ & Array steering vector at the transmitter&
${\mathbf{a}}_{{\mathrm{r}}}$ & Array steering vector at the receiver\\
$\bm{\theta}_{\mathrm{t}}$ & The AoD of multipath components&
$\bm{\theta}_{\mathrm{r}}$ & The AoA of multipath components\\
$\bm \beta$ & The channel gain of multipath components&
$\mathbf{G}$ & Effective channel matrix\\
$N_{\mathrm{s}}$ & Total number of applied multipath components&
$N_{\mathrm{s,a}}$ & Total number of applied orthogonal subchannels\\
$M$ & Modulation order of APM symbols&
$L$ & Total number of beam vector symbols \\
$m$ & Index of APM symbols&
$l$ & Index of beam vector symbols \\
$\mathbf{x}$ & The transmitted signal vector&
$\mathbf{y}$ & The received signal vector\\
$(\cdot)^{\mathrm{H}}$ & Complex conjugate transpose operator&
$\odot$ & Hadamard product\\
$\bm{\lambda}$ & Eigenvalues of $\mathbf{G}^\mathrm{H}\mathbf{G}$&
$\mathbf{U}$ & Eigenvectors of $\mathbf{G}^\mathrm{H}\mathbf{G}$\\
$\kappa$ & Index of subchannels&
$\bm\xi$&Weighting factors of subchannels\\
$\mathrm{chol}(\cdot)$ & Cholesky factorization of a matrix&
$\mathrm{Tr}(\cdot)$ & Trace of a matrix\\
$R\notdivides I$& $R$ is indivisible by $I$&
$|\cdot|$ & Magnitude of each complex element in the array\\
$\mathrm{E}[\cdot]$ & Expectation operator&
$Q(\cdot)$ & The $Q$-function\\
\hline
\end{tabular}
\vspace{-5mm}
\end{table}

\section{System model}
This section introduces the system model of the proposed MCA-SSM as shown in Figs. \ref{fig1}-\ref{fig2}. 

\subsection{Model assumptions}

\texttt{Assumption 1:}
In this paper, we consider a point-to-point connection with massive MIMO antenna arrays at both the Tx and the Rx.
The Tx array and the Rx array have $N_\mathrm{t}$ and $N_\mathrm{r}$ elements, respectively. The transmission from the Tx array to the Rx array can be expressed as
\begin{eqnarray}
\label{eq1}
{\bf{y}}= \sqrt{\frac{\rho}{N_{\mathrm{t}}}}\mathbf{H}\mathbf{x}+\mathbf{n}_{{\mathrm{a}}},
\end{eqnarray}
where 
$\rho$ is the SNR at the receiver,
$\bf{H}$ is the ${N_{\mathrm{r}}}\times{N_{\mathrm{t}}}$ channel matrix,
$\mathbf{x}$ is the ${N_{\mathrm{t}}}\times 1$ transmitted signal vector, 
$\mathbf{y}$ is the ${N_{\mathrm{r}}}\times 1$ received signal vector,
and 
$\mathbf{n_{\mathrm{a}}}\sim \mathcal{CN}(\mathbf{0},\mathbf{I})$ is the ${N_{\mathrm{r}}}\times 1$ noise vector. $\mathbf{H}$ is normalized to
$\mathrm{E}[\mathrm{Tr}({\mathbf{H}^H\mathbf{H}})]=N_{\mathrm{t}}N_{\mathrm{r}}$.

\texttt{Assumption 2:}
In this paper, perfect channel state information (CSI) is assumed to be known at both the Tx and the Rx following the setup of typical SSM systems \cite{Spatial, BIM, Millimeter,Generalized,Diversity,Polarized,YJiang,Adaptive, PY3, PY4}.

\texttt{Assumption 3:}
The narrowband sparse physical channel is adopted in this paper \cite{Spatial}. A link-level indoor channel is composed by 
$N_{{\mathrm{ts}}}$ MP components with different channel gains, AoDs, and AoAs. 
The channel matrix is given by \cite{goldsmith}
\begin{eqnarray}
\label{Hns}
{\bf{H}} = \sum\limits_{{n_{\mathrm{s}}} = 1}^{{N_{{\mathrm{ts}}}}} 
{
{\bm{\beta} (n_{\mathrm{s}})}
{\bf{\tilde H}}_{n_{\mathrm{s}}}
}, 
\end{eqnarray}
where 
$\bm{\beta}$ is the ${N_{\mathrm{ts}}}\times 1$
descending sorted channel gain vector. Its $n_{\mathrm{s}}$-th element, denoted by
$\bm{\beta}(n_{\mathrm{s}})$, is the channel gain from the Tx array to the Rx array via the $n_{\mathrm{s}}$-th MP component.
${\bf{\tilde H}}_{n_{\mathrm{s}}}$ is the normalized channel matrix from the Tx array to the Rx array via the $n_{\mathrm{s}}$-th MP component.
\begin{eqnarray}
\label{eq3}
{{{\bf{\tilde H}}}_{{n_{\mathrm{s}}}}} 
=
{\mathbf{a}}_{{\mathrm{r}},{n_\mathrm{s}}}^{\mathrm{H}}
{\mathbf{a}}_{{\mathrm{t}},{n_\mathrm{s}}},
\end{eqnarray}
${\mathbf{a}}_{{\mathrm{t}},{n_\mathrm{s}}}$
and
${\mathbf{a}}_{{\mathrm{r}},{n_\mathrm{s}}}$
are Tx and Rx array steering vectors, respectively, which are both row vectors  \cite{Spatial}.

\texttt{Assumption 4:}
For simplicity, assume that both the Tx and the Rx are equipped with uniform linear antenna arrays (ULAs) and all elements are omnidirectional and uni-polarized.
The $n_\mathrm{t}$-th element of ${\mathbf{a}}_{{\mathrm{t}}{n_\mathrm{s}}}$
and 
the $n_\mathrm{r}$-th element of ${\mathbf{a}}_{{\mathrm{r}}{n_\mathrm{s}}}$
are, respectively, given by 
\begin{eqnarray}
\label{77}
{\mathbf{a}}_{\mathrm{t},{n_\mathrm{s}}}(n_\mathrm{t}) = {e^{j\frac{2\pi}{\eta} \left(n_\mathrm{t}-\frac{1+N_\mathrm{t}}{2}\right){d_{{\mathrm{t}}}}\cos {\bm{\theta} _{\mathrm{t}}({n_{\mathrm{s}}})}}},
\end{eqnarray}
and 
\begin{eqnarray}
\label{77}
{\mathbf{a}}_{\mathrm{r},{n_\mathrm{s}}}(n_\mathrm{r}) = {e^{j\frac{2\pi}{\eta} \left(n_\mathrm{r}-\frac{1+N_\mathrm{r}}{2}\right){d_{{\mathrm{r}}}}\cos {\bm{\theta} _{\mathrm{r}}({n_{\mathrm{s}}})}}},
\end{eqnarray}
where $\eta$ denotes wavelength of the radio wave.
$\bm{\theta}_{\mathrm{t}}$ is the ${N_{\mathrm{ts}}}\times 1$ AoD vector. Its $n_{\mathrm{s}}$-th element, denoted by
$\bm{\theta} _{\mathrm{t}}(n_{\mathrm{s}})$, is the AoD of
the $n_{\mathrm{s}}$-th MP component. Similarly, $\bm{\theta}_{\mathrm{r}}$ is the AoA vector.
$d_\mathrm{t}$ and $d_\mathrm{r}$ denote adjacent spacing of Tx and Rx antenna elements, respectively.

\subsection{Transmitter}
In the proposed MCA-SSM system, information bits are divided into two streams to be transmitted at each instance. 
The total data rate is  $R=R_{\mathrm{mp}}+R_{\mathrm{sy}}=\log_2(ML)$ bits/symbol, where $M$ is the modulation order of the APM and $L$ is the number of candidate beam vector symbols.

To transmit the first stream, $s_{m}$ is selected from an APM constellation with $M$ candidate symbols, such as $M$-phase shift keying ($M$-PSK) and quadrature amplitude modulation ($M$-QAM). As such, the data rate of the first stream is $R_{\mathrm{sy}}=\log_2M$.

To transmit the second stream, a beam vector symbol out of $L$ MCA candidate beam vector symbols is selected to 
generate a beam pattern steering to $N_{\mathrm{s}}$ strongest MP components.
As such, the data rate of the second stream is $R_{\mathrm{mp}}=\log_2L$. 

The MCA mechanism for the second stream is illustrated in Fig. \ref{fig1}, and further elaborated as follows. 
When $N_{\mathrm{ts}}$ MP components are available in the wireless channel, 
Tx generates a beam pattern steering to 
 $N_{\mathrm{s}}$ candidate MP components out of $N_{\mathrm{ts}}$ MP components at any instance.
We choose MP components with $N_{\mathrm{s}}$ greatest channel gain to maximize the received power. 
For each selected MP component out of the $N_{\mathrm{s}}$ strongest MP components, a transmit phase shifter network is applied to steer a pencil beam to it based on the analog beamforming approach.
With \texttt{Assumption 2}, the Tx knows the number of MP components $N_{\mathrm{ts}}$, the gain of all selected MP components ${\bm\beta}(1:N_\mathrm{s})$, as well as their AoAs ${\bm\theta}_{\mathrm{t}}(1:N_\mathrm{s})$ and AoDs ${\bm\theta}_{\mathrm{r}}(1:N_\mathrm{s})$.
Then, the response of the $k$-th phase shifter network is given by
$\mathbf{a}^{\mathrm{H}}_{\mathrm{t},k}$, where $k\in{1,2,...,N_{\mathrm{s}}}$.
To aggregate the $N_{\mathrm{s}}$ strongest MP components, we use an MCA matrix $\mathbf{W}$ to weight and divide the power of the APM signal into $N_{\mathrm{s}}$
signals, and feed them into the $N_{\mathrm{s}}$ phase shifter networks.
If $R_{\mathrm{mp}}$ bits are transmitted in the second bit stream, $L=2^{R_{\mathrm{mp}}}$ rows of the MCA matrix $\mathbf{W}$ are candidates to be selected from. As such, 
$\mathbf{W}$ is an $L\times N_\mathrm{s}$ matrix. For the energy conservation law, $\mathbf{W}$ is normalized to  
$\|\mathbf{W}(l,:)\|^2=1$ throughout this paper.
The implementation of $\mathbf{W}$ is illustrated in Fig. \ref{fig2}. The design of $\mathbf{W}$ in this paper will be elaborated in Section IV.

Based on the above MCA mechanism, the overall response of the MCA matrix and all phase shifter networks at Tx is calculated by 
\begin{eqnarray}
\bm{\tau}_{{\mathrm t},l}=\sum_{k=1}^{N_\mathrm{s}}\mathbf{W}^*(l,k)\mathbf{a}^{\mathrm{H}}_{\mathrm{t},k}=[\mathbf{W}(l,:){\bf{A}}_{\mathrm{t}}(1:N_\mathrm{s},:)]^{\mathrm{H}},
\end{eqnarray}
where 
\begin{eqnarray}
{\bf{A}}_{\mathrm{t}}=\left[
\begin{smallmatrix}
{\mathbf{a}}_{\mathrm{t},1}
\\
{\mathbf{a}}_{\mathrm{t},2}
\\
...\\
{\mathbf{a}}_{\mathrm{t},N_\mathrm{ts}}
\end{smallmatrix}
\right].
\end{eqnarray}

Therefore, the transmitted signal at all Tx antenna elements, i.e., $\bf x$ in (\ref{eq1}), is 
\begin{eqnarray}
\label{eq7}
{\mathbf x}=
\bm{\tau}_{{\mathrm t},l}s_{m}=
\sum_{k=1}^{N_\mathrm{s}}\mathbf{W}^*(l,k)\mathbf{a}^{\mathrm{H}}_{\mathrm{t},k}s_{m}=[\mathbf{W}(l,:){\bf{A}}_{\mathrm{t}}(1:N_\mathrm{s},:)]^{\mathrm{H}}s_{m}.
\end{eqnarray}

\subsection{Receiver}

At the Rx, the received signal is computed by substituting (\ref{eq7}) into (\ref{eq1}).
\begin{eqnarray}
\mathbf{y} =\sqrt{\frac{\rho}{N_\mathrm{t}}}{\mathbf{H}}
[\mathbf{W}(l,:){\bf{A}}_{\mathrm{t}}(1:N_\mathrm{s},:)]^{\mathrm{H}}s_{m}
+\mathbf{n}_{\mathrm{a}}.
\end{eqnarray}

With \texttt{Assumption 3}, we have
\begin{eqnarray}
\label{eq8}
\mathbf{y} 
= \sqrt{\frac{\rho}{N_\mathrm{t}}}\sum\limits_{{n_{\mathrm{s}}} = 1}^{{N_{{\mathrm{ts}}}}} 
{\bm{\beta} (n_{\mathrm{s}})}{\mathbf{a}}_{{\mathrm{r}},{n_\mathrm{s}}}^{\mathrm{H}}
{\mathbf{a}}_{{\mathrm{t}},{n_\mathrm{s}}}
[\mathbf{W}(l,:){\bf{A}}_{\mathrm{t}}(1:N_\mathrm{s},:)]^{\mathrm{H}}s_{m}
+\mathbf{n}_{{\mathrm{a}}}
.
\end{eqnarray}

The received signal $\mathbf{y}$ is filtered by $N_{\mathrm{s}}$ Rx phase shifter networks and fed into $N_{\mathrm{s}}$ Rx RF chains. 
By assuming that Rx knows perfect CSI via channel estimation, each pair of Rx phase shifter network and Rx RF-chain is associated with one candidate MP component. 
To maximize the SNR at the RF chain, each Rx phase shifter network steers a pencil beam to its associated candidate MP component. 
That is, for the $k'$-th candidate MP component, the Rx phase shifter network is set as 
$
\mathbf{a}_{\mathrm{r},k'}$.

Denote the output signal of RF chains as $\mathbf{z}_0$.
Then, the output signal of the $k'$-th RF chain, $\mathbf{z}_0(k')$, is computed by 
multiplying both sides of (\ref{eq8}) by 
$
\mathbf{a}_{\mathrm{r},k'}$, i.e.,
\begin{eqnarray}
\begin{aligned}
\label{eq10}
\mathbf{z}_0(k')
&=\mathbf{a}_{\mathrm{r},k'}{{\mathbf{y}}} \\
&=\sqrt{\rho}
\underbrace{
\mathbf{a}_{\mathrm{r},k'}\sqrt{\frac{1}{N_\mathrm{t}}}\sum\limits_{{n_{\mathrm{s}}} = 1}^{{N_{{\mathrm{ts}}}}} 
{\bm{\beta} (n_{\mathrm{s}})}{\mathbf{a}}_{{\mathrm{r}},{n_\mathrm{s}}}^{\mathrm{H}}
{\mathbf{a}}_{{\mathrm{t}},{n_\mathrm{s}}}
\sum_{k=1}^{N_\mathrm{s}}\mathbf{W}^*(l,k)\mathbf{a}^{\mathrm{H}}_{\mathrm{t},k}
}_{\mathbf{\bar H}_0(k',l)}
s_{m}
+
\underbrace{\mathbf{a}_{\mathrm{r},k'}\mathbf{n}_{{\mathrm{a}}}}_{\mathbf{n}_{\mathrm{RF}}(k')}
\\
&=\sqrt{\rho}
{\mathbf{\bar H}_0(k',l)}
s_{m}
+
{\mathbf{n}_{\mathrm{RF}}(k')}
,
\end{aligned}
\end{eqnarray}
and then
\begin{eqnarray}
\label{eq10a}
\mathbf{z}_0=\sqrt{\rho}
{\mathbf{\bar H}_0(:,l)}
s_{m}
+
{\mathbf{n}_{\mathrm{RF}}}
.
\end{eqnarray}

To simplify the derivation of the optimal detection algorithm, an $N_{\mathrm{s}}\times L$ effective channel matrix
${\bf{\bar H}}_0$ is defined in (\ref{eq10}).
Following some algebraic manipulations,  $\mathbf{\bar H}_0$ is given by
\begin{eqnarray}
\label{eq11}
\mathbf{\bar H}_0=
\sqrt{\frac{1}{N_\mathrm{t}}}
{\bf{A}}_{\mathrm{r}}(1:N_\mathrm{s},:)
\sum\limits_{{n_{\mathrm{s}}} = 1}^{{N_{{\mathrm{ts}}}}} 
{\bm{\beta} (n_{\mathrm{s}})}{\mathbf{a}}_{{\mathrm{r}},{n_\mathrm{s}}}^{\mathrm{H}}
{\mathbf{a}}_{{\mathrm{t}},{n_\mathrm{s}}}
{\bf{A}}_{\mathrm{t}}^{\mathrm{H}}(1:N_\mathrm{s},:)\mathbf{W}^{\mathrm{H}}
,
\end{eqnarray}
where
\begin{eqnarray}
{\bf{A}}_{\mathrm{r}}=\left[
\begin{smallmatrix}
{\mathbf{a}}_{\mathrm{r},1}
\\
{\mathbf{a}}_{\mathrm{r},2}
\\
...\\
{\mathbf{a}}_{\mathrm{r},N_\mathrm{ts}}
\end{smallmatrix}
\right].
\end{eqnarray}

Moreover, since $\mathbf{n}_{\mathrm{RF}}$ is not white when $\mathbf{n}_{\mathrm{a}}$ is white, the covariance matrix of $\mathbf{n}_{\mathrm{RF}}$, denoted by 
$\mathbf{C}_{\mathrm{noise}}\triangleq \mathrm{E}[\mathbf{n}_{\mathrm{RF}}\mathbf{n}_{\mathrm{RF}}^\mathrm{H}]$ can be computed by
\begin{eqnarray}
\label{cndef}
\mathbf{C}_{\mathrm{noise}}
={\bf{A}}_{\mathrm{r}}(1:N_\mathrm{s},:){\bf{A}}_{\mathrm{r}}^\mathrm{H}(1:N_\mathrm{s},:),
\end{eqnarray}
which is not an identity matrix.

To whiten the noise, we multiply both sides of (\ref{eq10a}) by $\left({{\mathbf{B}}^{\mathrm{H}}}\right)^{-1}$, where matrix $\mathbf{B}$ is obtained by Cholesky factorization of ${{\bf{C}}_{{\mathrm{noise}}}}$, i.e.,
$\mathbf{B}=\mathrm{chol}(\mathbf{C}_{\mathrm{noise}})$ \cite{YJiang}. Then, we have
\begin{eqnarray}
\label{eq11e}
\begin{aligned}
\mathbf{z}
=\left({{\mathbf{B}}^{\mathrm{H}}}\right)^{-1}\mathbf{z}_0
=
\sqrt{\rho}
{\mathbf{G}}
\mathbf{W}^{\mathrm{H}}(l,:),
s_{m}
+
\underbrace{\left({{\mathbf{B}}^{\mathrm{H}}}\right)^{-1}{\mathbf{n}_{\mathrm{RF}}}}_{\mathcal{CN}(\mathbf{0},\mathbf{I})}
,
\end{aligned}
\end{eqnarray}
where $\mathbf{G}$ is an $N_\mathrm{s}\times N_\mathrm{s}$ matrix and computed by
\begin{eqnarray}
\label{eq11f}
\mathbf{G}
=\sqrt{\frac{1}{N_\mathrm{t}}}
\left({{\mathbf{B}}^{\mathrm{H}}}\right)^{-1}{\bf{A}}_{\mathrm{r}}(1:N_\mathrm{s},:)
\sum\limits_{{n_{\mathrm{s}}} = 1}^{{N_{{\mathrm{ts}}}}} 
{\bm{\beta} (n_{\mathrm{s}})}{\mathbf{a}}_{{\mathrm{r}},{n_\mathrm{s}}}^{\mathrm{H}}
{\mathbf{a}}_{{\mathrm{t}},{n_\mathrm{s}}}
{\bf{A}}_{\mathrm{t}}^{\mathrm{H}}(1:N_\mathrm{s},:).
\end{eqnarray}

Equation (\ref{eq11e}) is the same as the system model for conventional spatial modulation with 
equivalent channel matrix $\mathbf{GW^{\mathrm{H}}}$ and 
 additive white Gaussian noise $\left({{\mathbf{B}}^{\mathrm{H}}}\right)^{-1}{\mathbf{n}_{\mathrm{RF}}}$ \cite[Eq. (1)]{m6}. Then, applying a whitening filter and with the maximal likelihood (ML) detection, the optimal MCA-SSM detector is given by \cite[Eq. (4)]{optimalsm}

\begin{eqnarray}
\label{detectionalg}
[ \hat{l}, \hat{m}] = \mathop {\arg\min } \limits_{l,m} 
\left\{\left\|{\mathbf{z}}
-\sqrt{\rho}
\mathbf{GW^{\mathrm{H}}}(l,:)
s_{m}
\right\|\right\}.
\end{eqnarray}

\begin{remark}\textbf{State-of-the-art SSM systems are special realizations of the MCA-SSM:} It is worth noting that the proposed MCA-SSM system is a generalization of state-of-the-art SSM \cite{Spatial,BIM,Millimeter} and generalized SSM (GSSM) systems \cite{Generalized,YJiang}.
For 
$N_\mathrm{s}=L$, the proposed system is reduced to the SSM system \cite{Spatial,BIM,Millimeter} if $\mathbf{W}=\mathbf{I}$.
For $N_\mathrm{s}=5$ and $L=8$, the proposed system becomes the GSSM system \cite[Example 1]{YJiang} if  
\begin{eqnarray}
\mathbf{W}=
\frac{1}{\sqrt{2}}\left[
\begin{smallmatrix}
1&1&0&0&0\\
1&0&1&0&0\\
1&0&0&1&0\\
1&0&0&0&1\\
0&1&1&0&0\\
0&1&0&1&0\\
0&1&0&0&1\\
0&0&1&1&0\\
\end{smallmatrix}
\right].
\end{eqnarray}

\end{remark}

\section{Bit error probability analysis and problem formulation}

It is well known that analytical derivation of the exact ABEP with fading channels for complex modulation systems is challenging.
Therefore, a tight UUB on the ABEP for a given channel matrix $\mathbf{H}$ is derived as  \cite{upperbound}
\begin{eqnarray}
\label{eq14}
\mathrm{ABEP}_{\mathbf{H}} \leq 
\sum_{l_1=1}^{L}
\sum_{l_2=1}^{L}
\sum_{m_1=1}^{M}
\sum_{m_2=1}^{M}
\frac{
N(l_1,m_1,l_2,m_2)
P_{\mathrm{EP,\mathbf{H}}}(l_1,m_1,l_2,m_2)
}
{LM\log_2(LM)},
\end{eqnarray}
where $N(l_1,m_1,l_2,m_2)$ denotes
the number of bits in error when 
$l_1$ and $m_1$ are transmitted at Tx but $l_2$ and $m_2$ are detected at Rx.
$P_{\mathrm{EP,\mathbf{H}}}(l_1,m_1,l_2,m_2)$ is the pairwise error probability, which is defined as the probability that, when $l_1$ and $m_1$ are transmitted but $l_2$ and $m_2$ are detected assuming only candidates $(l_1,m_1)$ and $(l_2,m_2)$ can be selected at Tx. 
A closed-form expression of $P_{\mathrm{EP,\mathbf{H}}}(l_1,m_1,l_2,m_2)$ can be computed by
\begin{eqnarray}
P_{\mathrm{EP,\mathbf{H}}}(l_1,m_1,l_2,m_2)=Q\left(\sqrt{\frac{\rho J_{l_1,m_1,l_2,m_2}}{2}}\right),
\end{eqnarray}
where $J_{l_1,m_1,l_2,m_2}$ is the Euclidean distance (ED) between 
$\mathbf{GW^{\mathrm{H}}}(l_1,:)s_{m_1}$ and $\mathbf{GW^{\mathrm{H}}}(l_2,:)s_{m_2}$, which is computed by
\begin{eqnarray}
\label{Jdef}
\begin{aligned}
J_{l_1,m_1,l_2,m_2}&\triangleq&&
\left\|
\mathbf{GW^{\mathrm{H}}}(l_1,:)s_{m_1}
-
\mathbf{GW^{\mathrm{H}}}(l_2,:)s_{m_2}
\right\|^2\\
&=&&(
\mathbf{W}^\mathrm{H}(l_1,:)
s_{{m_1}}
-
\mathbf{W}^{\mathrm{H}}(l_2,:)
s_{m_2}
)^{\mathrm{H}}
\mathbf{G}^{\mathrm{H}}\mathbf{G}(
\mathbf{W}^\mathrm{H}(l_1,:)
s_{m_1}
-
\mathbf{W}^{\mathrm{H}}(l_2,:)
s_{m_2}
)
.
\end{aligned}
\end{eqnarray}

In this paper, we design the MCA mechanism by maximizing the 
minimum ED \cite{PY3,PY4}, which determines the ABEP in the high-SNR regime \cite{Diversity,GXu}.
Therefore, the optimization problem is formulated as
\begin{eqnarray}
\label{optimisation1}
\begin{aligned}
\mathrm{maximize}\ & \min\limits_
{
\begin{smallmatrix}
(l_1,m_1)\neq (l_2,m_2)\\
1\leq l_1\leq L, l\in \mathbb{N}^+\\
1\leq m_1\leq M, m\in \mathbb{N}^+\\
1\leq l_2\leq L, \hat{l}\in \mathbb{N}^+\\
1\leq m_2\leq M, \hat{m}\in \mathbb{N}^+
\end{smallmatrix}
}
J_{l_1,m_1,l_2,m_2},
\\
\mathrm{s.t.}\ & \|\mathbf{W}(l,:)\|=1, \forall l.
\end{aligned}
\end{eqnarray}

Apply eigendecomposition to $\mathbf{G}^\mathrm{H}\mathbf{G}$, and we have 
\begin{eqnarray}
\label{eig}
\mathbf{G}^\mathrm{H}\mathbf{G}= \mathbf{U} \mathbf{\Lambda} \mathbf{U} ^\mathrm{H},
\end{eqnarray}
where $\mathbf{U}$ is a uniform matrix, and $\mathbf{U}^\mathrm{H}\mathbf{U}=\mathbf{I}$. $\mathbf{\Lambda}$ is a  real diagonal matrix with descending sorted eigenvalues $\lambda_1,\lambda_2,...,\lambda_{N_\mathrm{s}}$.

By substituting (\ref{eig}) into (\ref{Jdef}) and defining $\mathbf{V}\triangleq(\mathbf{UW})^{\mathrm{H}}$ we have
\begin{eqnarray}
\begin{aligned}
J_{l_1,m_1,l_2,m_2}&=&&(
\mathbf{V}(:,l_1)
s_{{m_1}}
-
\mathbf{V}(:,l_2)
s_{m_2}
)^{\mathrm{H}}
\mathbf{\Lambda} 
(
\mathbf{V}(l_1,:)
s_{m_1}
-
\mathbf{V}(l_2,:)
s_{m_2}
)\\
&=&&\sum_{\kappa=1}^{N_\mathrm{s}}\lambda_\kappa
|\mathbf{V}(\kappa,l_1)
s_{m_1}
-
\mathbf{V}(\kappa,l_2)
s_{m_2}
|^2
\\
&=&&{\bm\lambda}
|\mathbf{V}(l_1,:)
s_{m_1}
-
\mathbf{V}(l_2,:)
s_{m_2}
|^2,
\end{aligned}
\end{eqnarray}
where ${\bm\lambda}\triangleq[\lambda_1,\lambda_2,...,\lambda_{N_\mathrm{s}}]$.
Under a given channel matrix, i.e., given $\lambda_\kappa$ and $\mathbf{U}$, we need to design the MCA matrix $\mathbf{W}$ to achieve a low ABEP calculated by (\ref{eq14}).

Noting that we can design $\mathbf{V}$ instead of $\mathbf{W}$, and obtain $\mathbf{W}$ by $\mathbf{W}=(\mathbf{V}\mathbf{U})^\mathrm{H}$, the optimization problem in (\ref{optimisation1}) is reformulated as
\begin{eqnarray}
\label{optimisation2}
\begin{aligned}
\mathrm{maximize}\ & \min\limits_
{
\begin{smallmatrix}
(l_1,m_1)\neq (l_2,m_2)\\
1\leq l_1\leq L, l\in \mathbb{N}^+\\
1\leq m_1\leq M, m\in \mathbb{N}^+\\
1\leq l_2\leq L, \hat{l}\in \mathbb{N}^+\\
1\leq m_2\leq M, \hat{m}\in \mathbb{N}^+
\end{smallmatrix}
}\{
{\bm\lambda}
|\mathbf{V}(l_1,:)
s_{m_1}
-
\mathbf{V}(l_2,:)
s_{m_2}
|^2\},
\\
\mathrm{s.t.}\ & \|\mathbf{V}(l,:)\|=1, \forall l.
\end{aligned}
\end{eqnarray}
Therein, the constraint $\|\mathbf{V}(l,:)\|=1, \forall l$ is applied to guarantee that $\|\mathbf{W}(l,:)\|=1, \forall l$, as $\mathbf{U}$ is a uniform matrix, and $\mathbf{W}=\mathbf{V}^{\mathrm{H}}\mathbf{U}$.

\section{Proposed multipath component aggregation mechanism}

In this section, we propose an approach to solve (\ref{optimisation2}). 
It is worth noting that achieving optimity of (\ref{optimisation2}) is challenging. 
Numerous techniques may be invoked for constructing the sub-optimal
matrix $\mathbf{V}$ for the classic spatial modulation (SM) systems  \cite{PY1,PY2,PY3,PY4}. 
Different from the SM system, the channel in (\ref{Hns}) is sparse and the eigenvalues $\bm\lambda$ differ from each other a lot.
Therefore, we apply $N_{\mathrm{s,a}}$ subchannels with the greatest eigenvalues to compose $\mathbf{V}$ as follows\footnote{
Besides tractability, another unexpected benefit is the simplification of the detection algorithm given by (\ref{detectionalg}), which can be reformulated as 
$[ \hat{l}, \hat{m}] = \mathop {\arg\min } \limits_{l,m} 
\left\{
\sum\limits_{\kappa=1}^{N_{\mathrm{s,a}}}
\left|{\mathbf{z}}(\kappa)
-
\sqrt{\rho\lambda_\kappa}{\bm\xi}(\kappa){\bm\upsilon}_l(\kappa)
s_{m}
\right|^2
\right\}$,
as
$
\left\|{\mathbf{z}}
-\sqrt{\rho}
\mathbf{GW^{\mathrm{H}}}(l,:)
s_{m}
\right\|^2=
\sum\limits_{\kappa=1}^{N_{\mathrm{s,a}}}
\left|{\mathbf{z}}(\kappa)
-
\sqrt{\rho\lambda_\kappa}{\bm\xi}(\kappa){\bm\upsilon}_l(\kappa)
s_{m}
\right|^2$. 
As such, i) the detector does not need to compute ${\mathbf{z}}((N_\mathrm{s,a}+1):N_\mathrm{s})$ and ii) the number of items in the summation is reduced from $N_{\mathrm{s}}$ to $N_{\mathrm{s,a}}$. Thus, the computational burden of the detection algorithm can be further reduced.
}. 
\begin{eqnarray}
\label{Vdesign}
\mathbf{V}(l,:)=({\bm \upsilon}_l\odot{\bm \xi})^\mathrm{H}, 
\end{eqnarray}
where $\odot$ denotes the Hadamard product, ${\bm \upsilon}_l$ is a vector that is associated with the $l$-th beam vector symbol, $\|{\bm \upsilon}_l\|=1, \forall l$, ${\bm \xi}$ is a real weighting factor that allocates power to subchannels with various eigenvalues, $\|{\bm \xi}\|=1$, and 
\begin{eqnarray}
\mathbf{V}(l,(N_{\mathrm{s,a}}+1):N_\mathrm{s})
={\bm \upsilon}_l((N_{\mathrm{s,a}}+1):N_\mathrm{s})
={\bm \xi}((N_{\mathrm{s,a}}+1):N_\mathrm{s})
=0, \forall l. 
\end{eqnarray}

\begin{example}
Learning from PSK, design of ${\bm \upsilon}_l$ can be achieved as follows with $N_{\mathrm{s,a}}=2$.
\begin{eqnarray}
{\bm \upsilon}_l=
\left[\cos\left(\frac{\pi}{4}+\frac{(l-1)\pi}{L}\right), \sin\left(\frac{\pi}{4}+\frac{(l-1)\pi}{L}\right),0,...,0\right].
\end{eqnarray}

For example, if $L=2$, we have
\begin{eqnarray}
\left\{
\begin{array}{l}
{\bm \upsilon}_1=\left[\frac{1}{\sqrt{2}},\frac{1}{\sqrt{2}},0,0\right],\\
{\bm \upsilon}_2=\left[-\frac{1}{\sqrt{2}},\frac{1}{\sqrt{2}},0,0\right].
\end{array}
\right.
\end{eqnarray}

If $L=4$, we have
\begin{eqnarray}
\label{upsilon4}
\left\{
\begin{array}{l}
{\bm \upsilon}_1=\left[\frac{1}{\sqrt{2}},\frac{1}{\sqrt{2}},0,0\right],\\
{\bm \upsilon}_2=\left[0,1,0,0\right],\\
{\bm \upsilon}_3=\left[-\frac{1}{\sqrt{2}},\frac{1}{\sqrt{2}},0,0\right],\\
{\bm \upsilon}_4=\left[1,0,0,0\right].
\end{array}
\right.
\end{eqnarray}
\end{example}

By substituting (\ref{Vdesign}) into (\ref{optimisation2}) and following some straightforward algebraic manipulations, the optimization problem can be rewritten as 

\begin{eqnarray}
\label{optimisation3}
\begin{aligned}
\mathrm{maximize} &\ \min\limits_
{
\begin{smallmatrix}
(l,m)\neq (\hat{l},\hat{m})\\
1\leq l_1\leq L, l\in \mathbb{N}^+\\
1\leq m_1\leq M, m\in \mathbb{N}^+\\
1\leq l_2\leq L, \hat{l}\in \mathbb{N}^+\\
1\leq m_2\leq M, \hat{m}\in \mathbb{N}^+
\end{smallmatrix}
}
\left\{
\sum_{\kappa=1}^{N_{\mathrm{s,a}}}{\bm \xi}^2(\kappa)\lambda_\kappa
|
{\bm \upsilon}_{l_1}(\kappa)
s_{m_1}
-
{\bm \upsilon}_{l_2}(\kappa)
s_{m_2}|^2\right\},
\\
\mathrm{s.t.} &\ \|{\bm \xi}\|=1.
\end{aligned}
\end{eqnarray}

The reformulated optimization problem in (\ref{optimisation3}) facilitates an analytic solution for the matrix $\mathbf{V}$, which is given by Theorem \ref{TheoremT1}.

\begin{theorem}\label{TheoremT1}\textbf{Analytic solution of the optimization problem (\ref{optimisation3}).}
The solution of (\ref{optimisation3}) is computed by
\begin{equation}
\label{eq38}
{\bm\xi}=\sqrt{
\frac{[1,{\iota}_{\mathrm{opt},2},...,{\iota}_{\mathrm{opt},N_{\mathrm{s,a}}},0,...,0]}{1+\sum_{\kappa=2}^{N_{\mathrm{s,a}}}{\iota}_{\mathrm{opt},\kappa}}},
\end{equation}
where 
${\bm\iota}_{\mathrm{opt}}=[1,{\iota}_{\mathrm{opt},2},...,{\iota}_{\mathrm{opt},N_{\mathrm{s,a}}}]$ is
one of ${\bm\iota}_{[n_{\mathrm{e},1},n_{\mathrm{e},2},...,n_{\mathrm{e},N_{\mathrm{s,a}}-1}]}$ that maximizes the minimal $J_{l_1,m_1,l_2,m_2}$ via traversing all possible $[n_{\mathrm{e},1},n_{\mathrm{e},2},...,n_{\mathrm{e},N_{\mathrm{s,a}}-1}]$ that satisfies $n_{\mathrm{e},1}<n_{\mathrm{e},2}<...<n_{\mathrm{e},N_{\mathrm{s,a}}-1}$.
${\bm\iota}_{[n_{\mathrm{e},1},n_{\mathrm{e},2},...,n_{\mathrm{e},N_{\mathrm{s,a}}-1}]}$
is the solution to the equation
\begin{eqnarray}
\label{eq33}
{\bm\epsilon}_{n_{\mathrm{e},1}}{\bm\iota}^\mathrm{T}={\bm\epsilon}_{n_{\mathrm{e},2}}{\bm\iota}^\mathrm{T}=...={\bm\epsilon}_{n_{\mathrm{e},n_{\mathrm{s,a}}}}{\bm\iota}^\mathrm{T}=...={\bm\epsilon}_{n_{\mathrm{e},N_{\mathrm{s,a}}-1}}{\bm\iota}^\mathrm{T},
\end{eqnarray}
${\bm\epsilon}_{n_{\mathrm{e},n_{\mathrm{s,a}}}}$ is an array with $N_{\mathrm{s,a}}$ elements, chosen from the set $\mathcal{D}$ without replacement. 
$\mathcal{D}$ is given by 
\begin{eqnarray}
\label{eq34}
\mathcal{D}=\mathcal{D}_0-
\{{\bm\epsilon}|{\bm\epsilon}\in\mathcal{D}_0, {\bm\epsilon}(\kappa)\geq {\bm\epsilon}'(\kappa),\forall\kappa, \exists{\bm\epsilon}'\in\mathcal{D}_0,{\bm\epsilon}'\neq{\bm\epsilon}\}.
\end{eqnarray}

For PSK signal constellation
\begin{eqnarray}
\begin{aligned}
\label{D1}
\mathcal{D}_0&=&&
\underbrace{\left\{
4\sin^2\left(\frac{\pi}{M}\right){\bm\lambda}({1:N_{\mathrm{s,a}}})\odot
|
{\bm \upsilon}_{l_1}(1:N_{\mathrm{s,a}})
|^2
\big|1\leq l_1\leq L, l_1\in \mathbb{N}^+
\right\}}_{\mathcal{D}_1}
\\
&&&\cup
\underbrace{\left\{
{\bm\lambda}({1:N_{\mathrm{s,a}}})\odot({\bm \upsilon}_{{l_1}}(1:N_{\mathrm{s,a}})\pm{\bm \upsilon}_{{l_2}}(1:N_{\mathrm{s,a}}))^2
\big|\begin{smallmatrix}
1\leq l_1\leq L, l_1\in \mathbb{N}^+\\
l_1<l_2\leq L, l_2\in \mathbb{N}^+
\end{smallmatrix}
\right\}}_{\mathcal{D}_2}.
\end{aligned}
\end{eqnarray}
For square QAM signal constellation
\begin{eqnarray}
\label{D2}
\begin{aligned}
\mathcal{D}_0&=&&
\underbrace{\left\{
\frac{6{\bm\lambda}({1:N_{\mathrm{s,a}}})\odot
|
{\bm \upsilon}_{l_1}(1:N_{\mathrm{s,a}})
|^2}{M-1}
\big|1\leq l_1\leq L, l_1\in \mathbb{N}^+
\right\}}_{\mathcal{D}_1}\\
&&&\cup
\underbrace{\left\{
{\bm\lambda}({1:N_{\mathrm{s,a}}})\odot|{\bm \upsilon}_{{l_1}}(1:N_{\mathrm{s,a}})s_{m_1}\pm{\bm \upsilon}_{{l_2}}(1:N_{\mathrm{s,a}})s_{m_2}|^2
\big|\begin{smallmatrix}
1\leq l_1\leq L, l_1\in \mathbb{N}^+\\
l_1<l_2\leq L, l_2\in \mathbb{N}^+\\
(s_{m_1},s_{m_2})\in\mathcal{S}_{\mathrm{m}}
\end{smallmatrix}
\right\}}_{\mathcal{D}_2},
\end{aligned}
\end{eqnarray}
where
\begin{eqnarray}
\label{smpsk}
\begin{aligned}
\mathcal{S}_{\mathrm{m}}=
\left\{(s_1,s_2)\left|
\begin{smallmatrix}
s_1=\sqrt{\frac{3}{2(M-1)}}(R_1+I_1), s_2=\sqrt{\frac{3}{2(M-1)}}(R_2+I_2)\\
R_1\in\{1,3,5,...,\sqrt{M}-1\}, 
I_1\in\{1,3,5,...,R_1\},
\\
R_2\in\{1,3,5,...,R_1\}, 
I_2\in\{1,3,5,...,I_1\},
\\
R_1\notdivides I_1, \forall (R_1\neq1),
R_2\notdivides I_2, \forall (R_2\neq1),\\
s_1\neq s_2,\forall(R_1,I_1)\neq(1,1)
\end{smallmatrix}
\right.\right\}.
\end{aligned}
\end{eqnarray} 
For rectangular QAM,
\begin{eqnarray}
\begin{aligned}
\mathcal{D}_0&=&&
\underbrace{\left\{
\frac{24{\bm\lambda}({1:N_{\mathrm{s,a}}})\odot
|
{\bm \upsilon}_{l_1}(1:N_{\mathrm{s,a}})
|^2}{5M-4}
\big|1\leq l_1\leq L, l_1\in \mathbb{N}^+
\right\}}_{\mathcal{D}_1}\\
&&&\cup
\underbrace{\left\{
{\bm\lambda}({1:N_{\mathrm{s,a}}})\odot|{\bm \upsilon}_{{l_1}}(1:N_{\mathrm{s,a}})s_{m_1}\pm{\bm \upsilon}_{{l_2}}(1:N_{\mathrm{s,a}})s_{m_2}|^2
\big|\begin{smallmatrix}
1\leq l_1\leq L, l_1\in \mathbb{N}^+\\
l_1<l_2\leq L, l_2\in \mathbb{N}^+\\
(s_{m_1},s_{m_2})\in\mathcal{S}_{\mathrm{m}}
\end{smallmatrix}
\right\}}_{\mathcal{D}_2},
\end{aligned}
\end{eqnarray}
where
\begin{eqnarray}
\label{smqam}
\begin{aligned}
\mathcal{S}_{\mathrm{m}}=
\left\{(s_1,s_2)\left|
\begin{smallmatrix}
s_1=\sqrt{\frac{6}{5M-4}}(R_1+I_1), s_2=\sqrt{\frac{6}{5M-4}}(R_2+I_2)\\
R_1\in\{1,3,5,...,\sqrt{2M}-1\}, 
I_1\in\{1,3,5,...,\min\{R_1,\sqrt{M/2}-1\}\},
\\
R_2\in\{1,3,5,...,R_1\}, 
I_2\in\{1,3,5,...,I_1\},
\\
R_1\notdivides I_1, \forall R_1\neq1,
R_2\notdivides I_2,\forall R_2\neq1,\\
s_1\neq s_2,\forall(R_1,I_1)\neq(1,1)
\end{smallmatrix}
\right.\right\}.
\end{aligned}
\end{eqnarray} 

Noting that $\mathcal{S}_{\mathrm{m}}$ needs to be deduced for MCA-SSM with QAM constellation diagrams.
Specifically, $\mathcal{S}_{\mathrm{m}}$ for $M=8,16,32,64$ are summarized in TABLE \ref{smtable}.
\end{theorem}
\begin{IEEEproof}
See Appendix \ref{ProveT1}.
\end{IEEEproof}

\begin{table}[h]
\label{smtable}
\centering
\caption{$\mathcal{S}_m$ for typical QAM constellations.}
\begin{tabular}{c|c}
\hline 
\textbf{Signal constellation} & $\mathcal{S}_m$\\
\hline
\hline
8-QAM& 
$\left\{\sqrt{\frac{1}{6}}(1+j,1+j),\sqrt{\frac{1}{6}}(1+j,3+j)\right\}$
\\
16-QAM&
$\left\{\sqrt{\frac{1}{10}}(1+j,1+j),\sqrt{\frac{1}{10}}(1+j,3+j)\right\}$
\\
32-QAM& 
$\left\{
\begin{smallmatrix}
\sqrt{\frac{1}{26}}(1+j,1+j),\sqrt{\frac{1}{26}}(1+j,3+j),
\sqrt{\frac{1}{26}}(1+j,5+j),\sqrt{\frac{1}{26}}(1+j,5+3j),\\
\sqrt{\frac{1}{26}}(1+j,7+j),\sqrt{\frac{1}{26}}(1+j,7+3j),
\sqrt{\frac{1}{26}}(3+j,5+j),\sqrt{\frac{1}{26}}(3+j,5+3j),\\
\sqrt{\frac{1}{26}}(3+j,7+j),\sqrt{\frac{1}{26}}(3+j,7+3j),
\sqrt{\frac{1}{26}}(5+j,5+3j),\sqrt{\frac{1}{26}}(5+j,7+j)\\
\sqrt{\frac{1}{26}}(5+j,7+3j),\sqrt{\frac{1}{26}}(5+3j,7+j)
\sqrt{\frac{1}{26}}(5+3j,7+3j),\sqrt{\frac{1}{26}}(7+j,7+3j)
\end{smallmatrix}
\right\}$
\\
64QAM &
$\left\{
\begin{smallmatrix}
\sqrt{\frac{1}{42}}(1+j,1+j),\sqrt{\frac{1}{42}}(1+j,3+j),
\sqrt{\frac{1}{42}}(1+j,5+j),\sqrt{\frac{1}{42}}(1+j,5+3j),\\
\sqrt{\frac{1}{42}}(1+j,7+j),\sqrt{\frac{1}{42}}(1+j,7+3j),
\sqrt{\frac{1}{42}}(1+j,7+5j),\sqrt{\frac{1}{42}}(3+j,5+j),\\
\sqrt{\frac{1}{42}}(3+j,5+3j),\sqrt{\frac{1}{42}}(3+j,7+j),
\sqrt{\frac{1}{42}}(3+j,7+3j),\sqrt{\frac{1}{42}}(3+j,7+5j),\\
\sqrt{\frac{1}{42}}(5+j,5+3j),\sqrt{\frac{1}{42}}(5+j,7+j),
\sqrt{\frac{1}{42}}(5+j,7+3j),\sqrt{\frac{1}{42}}(5+j,7+5j),\\
\sqrt{\frac{1}{42}}(5+3j,7+j),\sqrt{\frac{1}{42}}(5+3j,7+3j),
\sqrt{\frac{1}{42}}(5+3j,7+5j),\sqrt{\frac{1}{42}}(7+j,7+3j),\\
\sqrt{\frac{1}{42}}(7+j,7+5j),\sqrt{\frac{1}{42}}(7+3j,7+5j)
\end{smallmatrix}
\right\}$
\\\hline 
\end{tabular}
%
\vspace{0.1in}
\centering
\caption{Channel parameters applied in Examples 2 and 3.}
\label{table_exampparam}
\begin{tabular}{c c c c}
\hline 
MP\#& $\beta$ & $\theta_\mathrm{t}$ & $\theta_\mathrm{r}$\\
\hline
\hline
1& 0.9356&2&2.0\\
2&-0.2807&2.05&1.6\\
3& 0.1871&1.2&2.4\\
4&-0.0936&3&2.45\\
5& 0.0468&0.4&2.8\\
\hline
\end{tabular}
\end{table}

\section{Case study}

\subsection{Cases under consideration}

To clearly explain how to apply Theorem \ref{TheoremT1} in the proposed MCA-SSM system, 
we consider Example 2 and Example 3 for 16-PSK and 16-QAM, respectively, with $N_{\mathrm{s,a}}=2$, $L=4$, and $M=16$. Then, ${\bm \upsilon}_{l}$ is given by (\ref{upsilon4}).
For other values of $N_{\mathrm{s}}$, $N_{\mathrm{s,a}}$, $L=4$, and $M$, Theorem \ref{TheoremT1} can be applied using a similar approach.

Channel parameters for both examples are listed in TABLE \ref{table_exampparam}.
We further assign $N_{\mathrm{s}}=4$. With these channel parameters, we can obtain that ${\bm\lambda}=[410.05,9.84,1.41,0.05]$ and 
\begin{eqnarray}
\mathbf{U}=\left[
\begin{smallmatrix}
    0.0697&   -0.6370&    0.0568&    0.7656\\
    0.0275&    0.7688&   -0.0187&    0.6386\\
   -0.3857&   -0.0558&   -0.9192&    0.0569\\
   -0.9196&   -0.0019&    0.3892&    0.0533\\
\end{smallmatrix}
\right]=\left[
\begin{smallmatrix}
\mathbf{u}_1\\
\mathbf{u}_2\\
\mathbf{u}_3\\
\mathbf{u}_4
\end{smallmatrix}
\right].
\end{eqnarray}

\begin{example}\label{example2}\textbf{16-PSK signal constellation.}

Firstly, as $|
{\bm \upsilon}_{1}(1)
|^2=|
{\bm \upsilon}_{1}(2)
|^2=|
{\bm \upsilon}_{3}(1)
|^2=|
{\bm \upsilon}_{3}(2)
|^2=1/2$, 
$|
{\bm \upsilon}_{2}(1)
|^2=|
{\bm \upsilon}_{4}(2)
|^2=1$,
$|
{\bm \upsilon}_{2}(2)
|^2=|
{\bm \upsilon}_{4}(1)
|^2=0$,
and
$\sin^2\left(\frac{\pi}{16}\right)=\frac{1}{4}\left(2-\sqrt{2+\sqrt{2}}\right)$
 we have
\begin{eqnarray}
\label{eq46}
\mathcal{D}_1=
\left\{
\left(1-\sqrt{\frac{2+\sqrt{2}}{4}}\right)
[\lambda_2,\lambda_1],
\left(2-\sqrt{2+\sqrt{2}}\right)
[\lambda_1,0],
\left(2-\sqrt{2+\sqrt{2}}\right)
[0,\lambda_2]
\right\}.
\end{eqnarray}

Secondly,
\begin{eqnarray}
\label{eq47}
\mathcal{D}_2=
\left\{
\left[4\lambda_{1},0\right],\left[0,4\lambda_{2}\right],
\left[\left(1-\frac{1}{\sqrt{2}}\right)^2\lambda_{1},\frac{1}{2}\lambda_{2}\right],
\left[\frac{1}{2}\lambda_{1},\left(1-\frac{1}{\sqrt{2}}\right)^2\lambda_{2}\right]
\right\},
\end{eqnarray}
as
\begin{eqnarray}
&&\left\{
{\bm\lambda}({1:N_{\mathrm{s,a}}})\odot({\bm \upsilon}_{{l_1}}(1:N_{\mathrm{s,a}})-{\bm \upsilon}_{{l_2}}(1:N_{\mathrm{s,a}}))^2
\big|\begin{smallmatrix}
1\leq l_1\leq L, l_1\in \mathbb{N}^+\\
l_1<l_2\leq L, l_2\in \mathbb{N}^+
\end{smallmatrix}
\right\}\nonumber
\\&&=\left\{
\left[4\lambda_{1},0\right],
\left[\left(1-\frac{1}{\sqrt{2}}\right)^2\lambda_{1},\frac{1}{2}\lambda_{2}\right],
\left[\frac{1}{2}\lambda_{1},\left(1-\frac{1}{\sqrt{2}}\right)^2\lambda_{2}\right]
\right\},
\end{eqnarray}
and
\begin{eqnarray}
&&\left\{
{\bm\lambda}({1:N_{\mathrm{s,a}}})\odot({\bm \upsilon}_{{l_1}}(1:N_{\mathrm{s,a}})+{\bm \upsilon}_{{l_2}}(1:N_{\mathrm{s,a}}))^2
\big|\begin{smallmatrix}
1\leq l_1\leq L, l_1\in \mathbb{N}^+\\
l_1<l_2\leq L, l_2\in \mathbb{N}^+
\end{smallmatrix}
\right\}\nonumber
\\&&=\left\{
\left[0,4\lambda_{2}\right],
\left[\left(1-\frac{1}{\sqrt{2}}\right)^2\lambda_{1},\frac{1}{2}\lambda_{2}\right],
\left[\frac{1}{2}\lambda_{1},\left(1-\frac{1}{\sqrt{2}}\right)^2\lambda_{2}\right]
\right\}.
\end{eqnarray}

Then, combining (\ref{eq46}) and (\ref{eq47}),
and with
$\left(2-\sqrt{2+\sqrt{2}}\right)<4$,
$\left(2-\sqrt{2+\sqrt{2}}\right)<\left(1-\frac{1}{\sqrt{2}}\right)^2$,
$\left(2-\sqrt{2+\sqrt{2}}\right)<\frac{1}{2}$,
 we have
\begin{eqnarray}
\mathcal{D}=
\left\{
\begin{array}{l}
\left(1-\sqrt{\frac{2+\sqrt{2}}{4}}\right)
[\lambda_1,\lambda_2],\\
\left(2-\sqrt{2+\sqrt{2}}\right)
[\lambda_1,0],\\
\left(2-\sqrt{2+\sqrt{2}}\right)
[0,\lambda_2],
\end{array}
\right\}
\end{eqnarray}

According to Theorem \ref{TheoremT1}, the solution of (\ref{optimisation3}) is one of the following equations.
\begin{eqnarray}
\left\{
\begin{array}{l}
\left(1-\sqrt{\frac{2+\sqrt{2}}{4}}\right)(\lambda_2\iota_2+\lambda_1)=
\left(2-\sqrt{2+\sqrt{2}}\right)\lambda_{1},
\\
\left(2-\sqrt{2+\sqrt{2}}\right)\lambda_{2}\iota_2=\left(2-\sqrt{2+\sqrt{2}}\right)\lambda_{1}.
\end{array}
\right.
\end{eqnarray}

Therefore, the only possible solution is
\begin{eqnarray}
\label{optimaliotaPSK}
\iota_{\mathrm{opt},2}
=
\frac{\lambda_1}{\lambda_2}
\end{eqnarray}

By substituting (\ref{optimaliotaPSK}) into (\ref{eq38}) and with some straightforward derivations, we obtain
\begin{eqnarray}
{\bm\xi}^2=\left[\frac{\lambda_2}{\lambda_1+\lambda_2}, \frac{\lambda_1}{\lambda_1+\lambda_2}\right],
\end{eqnarray}
\begin{eqnarray}
\mathbf{V}=\left[
\begin{array}{cccc}
\frac{1}{\sqrt{2}}\sqrt{\frac{\lambda_2}{\lambda_1+\lambda_2}} &\frac{1}{\sqrt{2}}\sqrt{\frac{\lambda_1}{\lambda_1+\lambda_2}}&0&0\\
0&\sqrt{\frac{\lambda_1}{\lambda_1+\lambda_2}}&0&0\\
-\frac{1}{\sqrt{2}}\sqrt{\frac{\lambda_2}{\lambda_1+\lambda_2}}&\frac{1}{\sqrt{2}}\sqrt{\frac{\lambda_1}{\lambda_1+\lambda_2}}&0&0\\
\sqrt{\frac{\lambda_2}{\lambda_1+\lambda_2}}&0&0&0
\end{array}
\right],
\end{eqnarray}
and thus
\begin{eqnarray}
\label{Wexample2}
\mathbf{W}=\left[
\begin{array}{c}
\frac{1}{\sqrt{2}}\sqrt{\frac{\lambda_2}{\lambda_1+\lambda_2}} \mathbf{u}_1+\frac{1}{\sqrt{2}}\sqrt{\frac{\lambda_1}{\lambda_1+\lambda_2}}\mathbf{u}_2\\
\sqrt{\frac{\lambda_1}{\lambda_1+\lambda_2}}\mathbf{u}_2\\
-\frac{1}{\sqrt{2}}\sqrt{\frac{\lambda_2}{\lambda_1+\lambda_2}} \mathbf{u}_1+\frac{1}{\sqrt{2}}\sqrt{\frac{\lambda_1}{\lambda_1+\lambda_2}}\mathbf{u}_2\\
\sqrt{\frac{\lambda_2}{\lambda_1+\lambda_2}} \mathbf{u}_1
\end{array}
\right].
\end{eqnarray}

\end{example}

\begin{example}\label{example3}\textbf{$N_{\mathrm{s,a}}=2$, $L=4$, 16-QAM signal constellation.}

For the 16-QAM signal constellation, 
\begin{eqnarray}
\mathcal{D}_1=
\left\{
\frac{1}{5}
[\lambda_1,\lambda_2],
\frac{2}{5}
[\lambda_1,0],
\frac{2}{5}
[0,\lambda_2]
\right\}.
\end{eqnarray}

Then, 
\begin{eqnarray}
\mathcal{S}_m=\left\{\sqrt{\frac{1}{10}}(1+j,1+j),\sqrt{\frac{1}{10}}(1+j,3+j)\right\},
\end{eqnarray}
and we have
\begin{eqnarray}
\mathcal{D}_2
=
\left\{
\begin{array}{l}
\frac{2}{5}
[\lambda_1,0],
\frac{2}{5}
[0,\lambda_2],
\frac{1}{5}\left[\left(1-\frac{1}{\sqrt{2}}\right)^2\lambda_1,\frac{1}{2}\lambda_2\right],
\frac{1}{5}\left[\frac{1}{2}\lambda_1,\left(1-\frac{1}{\sqrt{2}}\right)^2\lambda_2\right],\\
\frac{1}{10}\left[\frac{|1+3j|^2}{2}\lambda_1,\left|(1+j)-\frac{1+3j}{\sqrt{2}}\right|^2\lambda_2\right],
\frac{1}{10}\left[\left|(1+j)-\frac{1+3j}{\sqrt{2}}\right|^2\lambda_1,\frac{|1+3j|^2}{2}\lambda_2\right],\\
\frac{1}{10}\left[\frac{|1+1j|^2}{2}\lambda_1,\left|(1+3j)-\frac{1+j}{\sqrt{2}}\right|^2\lambda_2\right],
\frac{1}{10}\left[\left|(1+3j)-\frac{1+j}{\sqrt{2}}\right|^2\lambda_1,\frac{|1+j|^2}{2}\lambda_2\right]
\end{array}
\right\}.
\end{eqnarray}

Because 
$\left|(1+3j)-\frac{1+j}{\sqrt{2}}\right|^2>1$, $\frac{|1+j|^2}{2}=1$, $\left|(1+j)-\frac{1+3j}{\sqrt{2}}\right|^2>1$, $\frac{|1+3j|^2}{2}>1$, $\left(1-\frac{1}{\sqrt{2}}\right)^2<1$, 
we obtain
\begin{eqnarray}
\mathcal{D}=
\left\{
\begin{array}{l}
\frac{2}{5}[\lambda_1,0],
\frac{2}{5}
[0,\lambda_2],
\frac{1}{5}\left[\left(1-\frac{1}{\sqrt{2}}\right)^2\lambda_1,\frac{1}{2}\lambda_2\right],
\frac{1}{5}\left[\frac{1}{2}\lambda_1,\left(1-\frac{1}{\sqrt{2}}\right)^2\lambda_2\right]
\end{array}
\right\}.
\end{eqnarray}

Therefore, according to Theorem \ref{TheoremT1}, the solution of (\ref{optimisation3}) is one of the following equations.
\begin{eqnarray}
\left\{
\begin{array}{l}
\frac{2}{5}\lambda_2\iota_2=\frac{2}{5}\lambda_1,\\
\frac{2}{5}\lambda_2\iota_2=\frac{1}{5}\left[\left(1-\frac{1}{\sqrt{2}}\right)^2\lambda_1+\frac{1}{2}\lambda_2\iota_2\right],\\
\frac{2}{5}\lambda_2\iota_2=\frac{1}{5}\left[\frac{1}{2}\lambda_1+\left(1-\frac{1}{\sqrt{2}}\right)^2\lambda_2\iota_2\right],\\
\frac{2}{5}\lambda_1=\frac{1}{5}\left[\left(1-\frac{1}{\sqrt{2}}\right)^2\lambda_1+\frac{1}{2}\lambda_2\iota_2\right],\\
\frac{2}{5}\lambda_1=\frac{1}{5}\left[\frac{1}{2}\lambda_1+\left(1-\frac{1}{\sqrt{2}}\right)^2\lambda_2\iota_2\right],\\
\frac{1}{5}\left[\left(1-\frac{1}{\sqrt{2}}\right)^2\lambda_1+\frac{1}{2}\lambda_2\iota_2\right]
=\frac{1}{5}\left[\frac{1}{2}\lambda_1+\left(1-\frac{1}{\sqrt{2}}\right)^2\lambda_2\iota_2\right].
\end{array}
\right.
\end{eqnarray}
Then, the set of possible solutions are given by
\begin{eqnarray}
\label{optimaliotaQAM}
\iota_{\mathrm{opt},2}
\in\left\{
\frac{(\sqrt{2}-1)^2}{3}\frac{\lambda_1}{\lambda_2},
\frac{1}{1+2\sqrt{2}}\frac{\lambda_1}{\lambda_2},
\frac{\lambda_1}{\lambda_2},
({1+2\sqrt{2}})\frac{\lambda_1}{\lambda_2},
\frac{3}{(\sqrt{2}-1)^2}\frac{\lambda_1}{\lambda_2}
\right\}.
\end{eqnarray}

By investigating the five candidates, it is not difficult to figure out that $\iota_{\mathrm{opt},2}
=
\frac{(\sqrt{2}-1)^2}{3}
\frac{\lambda_1}{\lambda_2}$ maximizes $\min\{J_{l_1,m_1,l_2,m_2}\}$, as shown in TABLE. \ref{t_example3a}, and therefore it is the solution of the optimization problem (\ref{optimisation3}).

\begin{table}[h]
\centering
\caption{Candidate solutions of $\iota_{2}$ for Example 3 and their achievable minimal ED, $\min\{J_{l_1,m_1,l_2,m_2}\}$.}
\label{t_example3a}
\begin{tabular}{c|c}
\hline 
\bf{Candidates of} $\iota_{\mathrm{opt},2}$& \bf{Achievable} $\min\{J_{l_1,m_1,l_2,m_2}\}$  \\ 
\hline
\hline
$\frac{3\lambda_1}{(\sqrt{2}-1)^2\lambda_2}$
&
0.4497
\\
$\frac{(1+2\sqrt{2})\lambda_1}{\lambda_2}$
&
0.8466
\\
$\frac{\lambda_1}{\lambda_2}$
&
2.2520
\\
$\frac{\lambda_1}{(1+2\sqrt{2})\lambda_2}$
&
2.9869
\\
$\frac{(\sqrt{2}-1)^2\lambda_1}{3\lambda_2}$ 
&
5.5460
\\
\hline
\end{tabular}
\end{table}

By substituting 
$\iota_{\mathrm{opt},2}
=
\frac{(\sqrt{2}-1)^2}{3}
\frac{\lambda_1}{\lambda_2}$
 into (\ref{eq38}) and following some straightforward derivations, we obtain
\begin{eqnarray}
{\bm\xi}^2=\left[\frac{3\lambda_2}{(\sqrt{2}-1)^2\lambda_1+3\lambda_2}, 
\frac{(\sqrt{2}-1)^2\lambda_1}{(\sqrt{2}-1)^2\lambda_1+3\lambda_2}\right],
\end{eqnarray}
\begin{eqnarray}
\mathbf{V}=\left[
\begin{array}{cccc}
\frac{1}{\sqrt{2}}\sqrt{\frac{3\lambda_2}{(\sqrt{2}-1)^2\lambda_1+3\lambda_2}} &
\frac{1}{\sqrt{2}}\sqrt{\frac{(\sqrt{2}-1)^2\lambda_1}{(\sqrt{2}-1)^2\lambda_1+3\lambda_2}}
&0&0\\
0&\sqrt{\frac{(\sqrt{2}-1)^2\lambda_1}{(\sqrt{2}-1)^2\lambda_1+3\lambda_2}}&0&0\\
-\frac{1}{\sqrt{2}}\sqrt{\frac{3\lambda_2}{(\sqrt{2}-1)^2\lambda_1+3\lambda_2}}&
\frac{1}{\sqrt{2}}\sqrt{\frac{(\sqrt{2}-1)^2\lambda_1}{(\sqrt{2}-1)^2\lambda_1+3\lambda_2}}&
0&0\\
\sqrt{\frac{3\lambda_2}{(\sqrt{2}-1)^2\lambda_1+3\lambda_2}}&0&0&0
\end{array}
\right],
\end{eqnarray}
and thus
\begin{eqnarray}
\label{Wexample3}
\mathbf{W}=\left[
\begin{array}{c}
\frac{1}{\sqrt{2}}\sqrt{\frac{3\lambda_2}{(\sqrt{2}-1)^2\lambda_1+3\lambda_2}} \mathbf{u}_1
+\frac{1}{\sqrt{2}}\sqrt{\frac{(\sqrt{2}-1)^2\lambda_1}{(\sqrt{2}-1)^2\lambda_1+3\lambda_2}}\mathbf{u}_2\\
\sqrt{\frac{(\sqrt{2}-1)^2\lambda_1}{(\sqrt{2}-1)^2\lambda_1+3\lambda_2}}\mathbf{u}_2\\
-\frac{1}{\sqrt{2}}\sqrt{\frac{3\lambda_2}{(\sqrt{2}-1)^2\lambda_1+3\lambda_2}} \mathbf{u}_1
+\frac{1}{\sqrt{2}}\sqrt{\frac{(\sqrt{2}-1)^2\lambda_1}{(\sqrt{2}-1)^2\lambda_1+3\lambda_2}}\mathbf{u}_2\\
\sqrt{\frac{3\lambda_2}{(\sqrt{2}-1)^2\lambda_1+3\lambda_2}} \mathbf{u}_1
\end{array}
\right].
\end{eqnarray}

\end{example}

\subsection{Numerical results and discussions}

\begin{figure}[h]
\centering
\subfigure[ED versus $\iota$ for Example 2]{\includegraphics[scale=0.5]{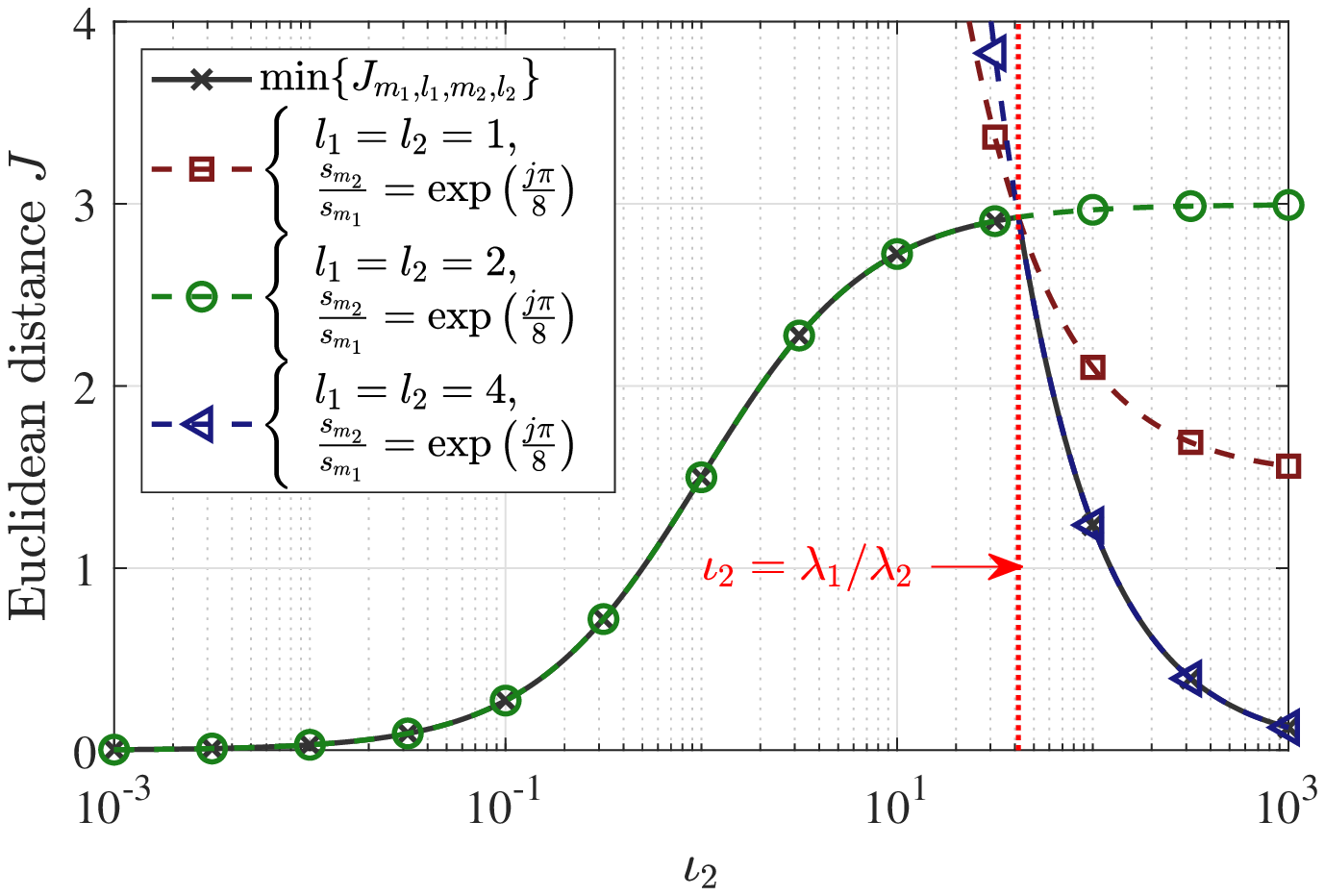}}
\subfigure[ED versus $\iota$ for Example 3]{\includegraphics[scale=0.5]{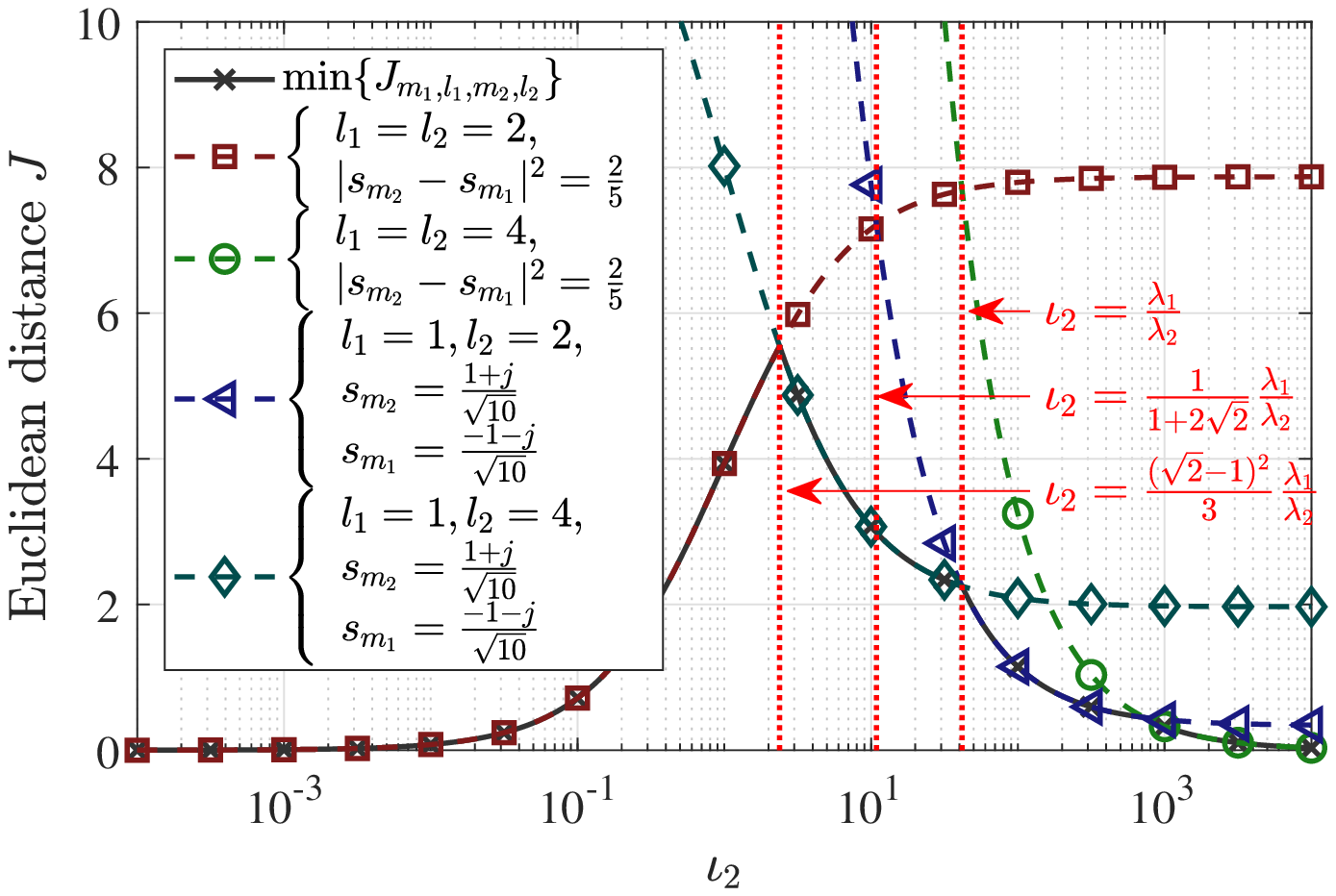}}
\caption{Numerical validation of minimum ED maximization.}\label{EDvalidation}
\end{figure}

To validate Theorem \ref{TheoremT1}, the ED result versus $\iota_2$ is illustrated in Fig. \ref{EDvalidation}(a) and Fig. \ref{EDvalidation}(b) for Example 2 and Example 3, respectively. In Fig. \ref{EDvalidation}(a), it can be seen that $\iota_{\mathrm{opt},2}=\frac{\lambda_1}{\lambda_2}$ maximizes the minimal ED, i.e., $\min_{J_{l_1,m_1,l_2,m_2}}$ in (\ref{optimisation1}), and in Fig. \ref{EDvalidation}(b), $\iota_{\mathrm{opt},2}
=
\frac{(\sqrt{2}-1)^2}{3}
\frac{\lambda_1}{\lambda_2}$ maximizes the minimal ED.

\begin{figure}[h]
\centering
\subfigure[Conventional SSM]{\includegraphics[scale=0.5]{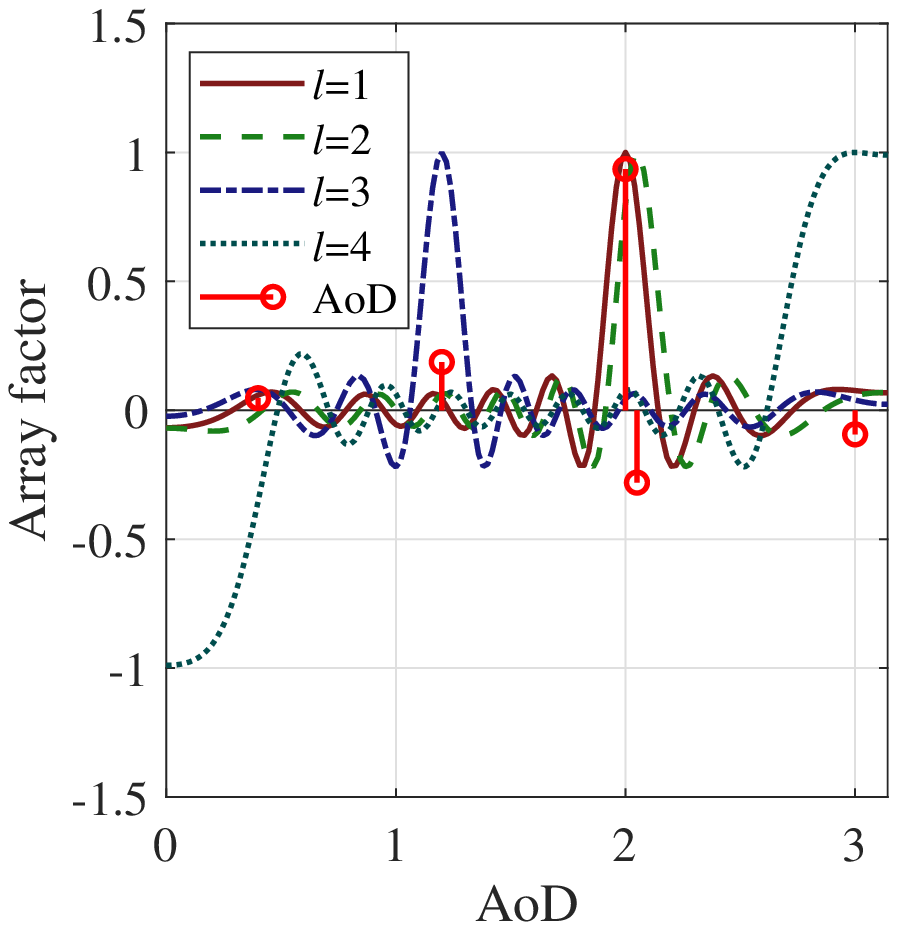}}
\subfigure[MCA-SSM with 16-PSK]{\includegraphics[scale=0.5]{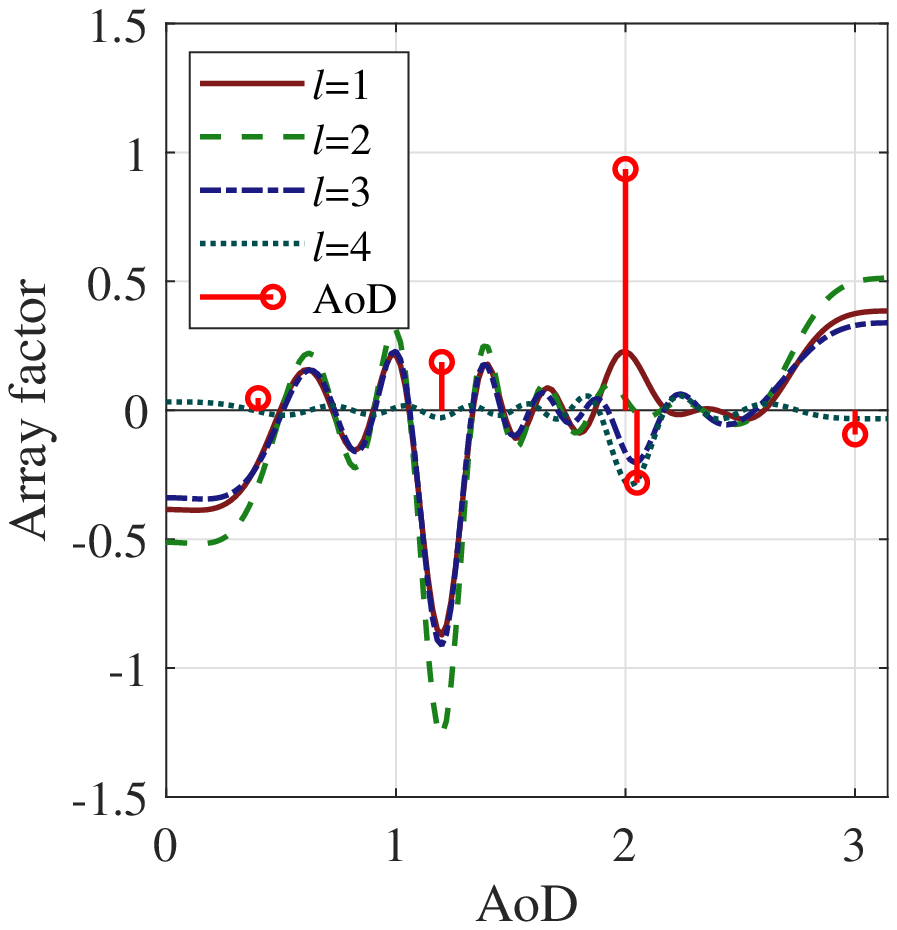}}
\subfigure[MCA-SSM with 16-QAM]{\includegraphics[scale=0.5]{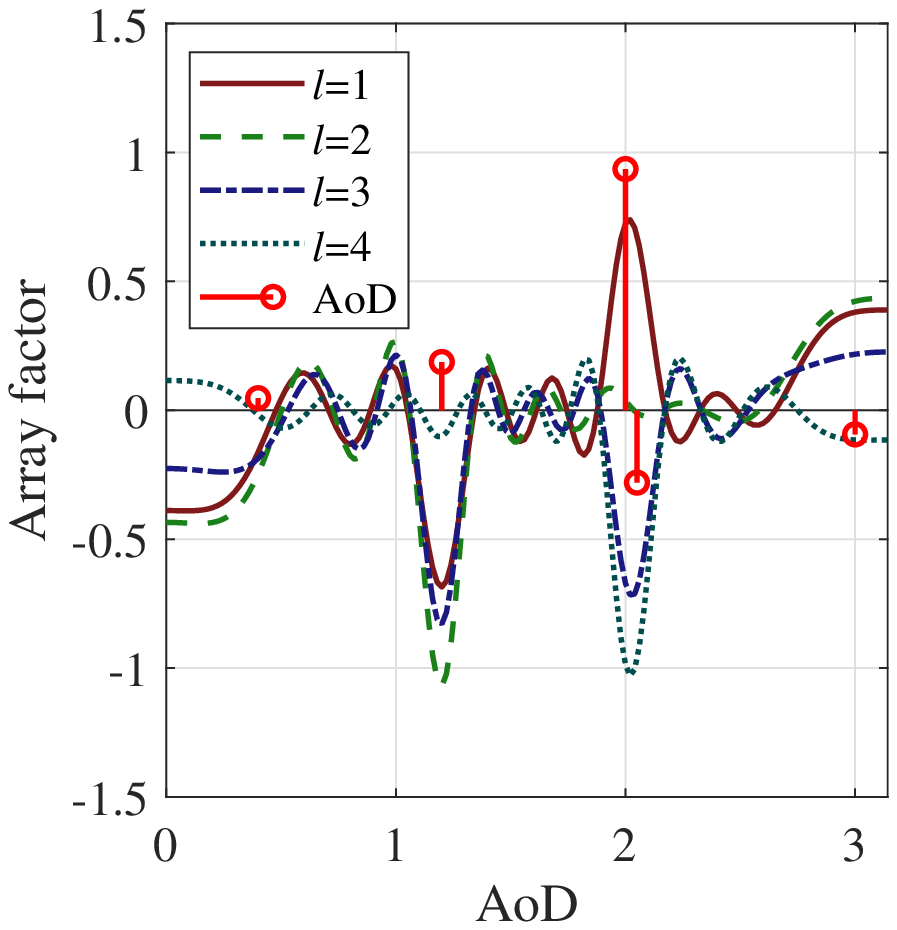}}
\caption{Transmit array factors for the conventional SSM system and proposed MCA-SSM system.}\label{fig6}
\end{figure}

Now, we take a deeper look into the transmit array factors for the conventional SSM system and the proposed MCA-SSM, illustrated in Fig. \ref{fig6}. For the conventional SSM, $\mathbf{W}=\mathbf{I}$, and the beam is steered to only one MP component for each value of $l$. For the MCA-SSM with 16-PSK and 16-QAM, $\mathbf{W}$ is given by (\ref{Wexample2}) and (\ref{Wexample3}), respectively, and MP components are aggregated for each value of $l$. This means that most of the four strongest MP components are steered for each value of $l$.

The task of the receiver is to detect the transmit array factor out of the illustrated four array factors
 while decoding $\hat{l}$.
Unfortunately,  the transmit array factors for $l=1$ and $l=2$ in the conventional SSM system are similar as the value of 
${\bm\theta}_{\mathrm{t}}(1)$ is close to the value of ${\bm\theta}_{\mathrm{t}}(2)$. 
Under this situation, ED for $(l_1, l_2)=(1,2)$ is small and leads to a large pairwise error performance $P_{\mathrm{EP,\mathbf{H}}}(1,m_1,2,m_1),\forall m_1\in\{1,2,...,M\}$.
In contrast, the shapes of four array factors for the proposed MCA-SSM are quite different and easy to distinguish at the detector.

\begin{figure}[h]
\centering
\subfigure[ABEP versus SNR for Example 2]{\includegraphics[scale=0.5]{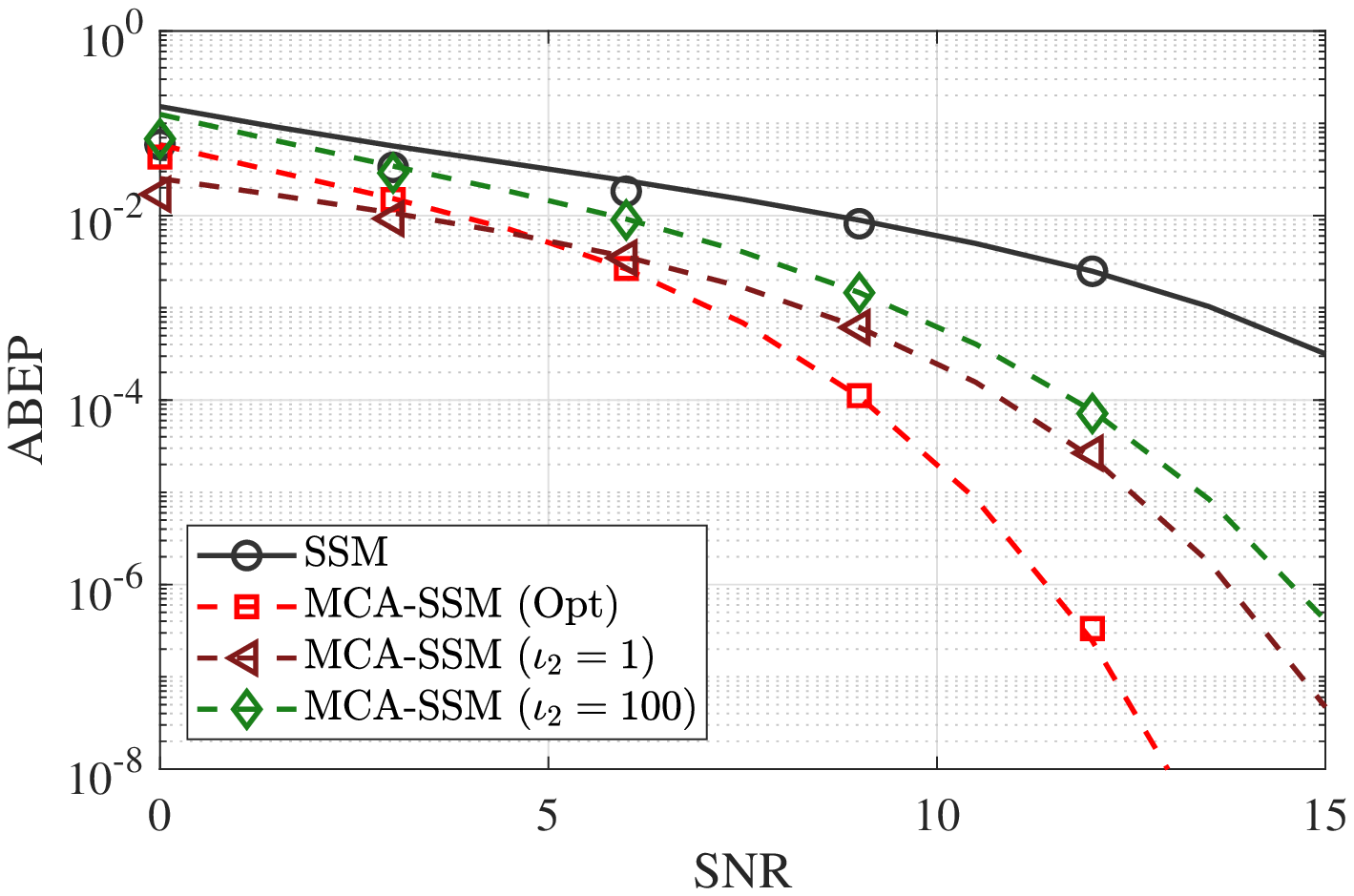}}
\subfigure[ABEP versus SNR for Example 3]{\includegraphics[scale=0.5]{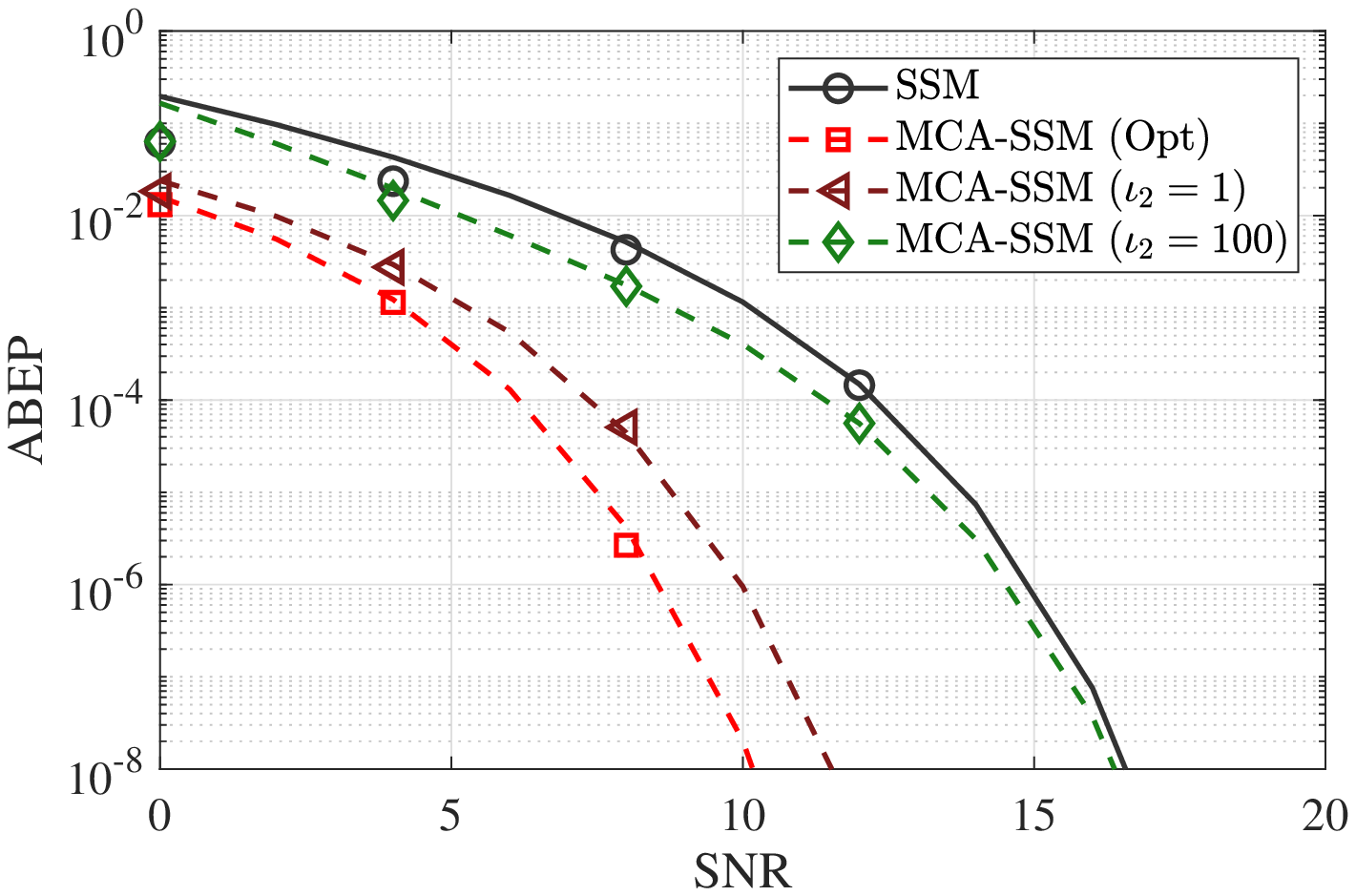}}
\subfigure[ABEP versus $\iota$]{\includegraphics[scale=0.5]{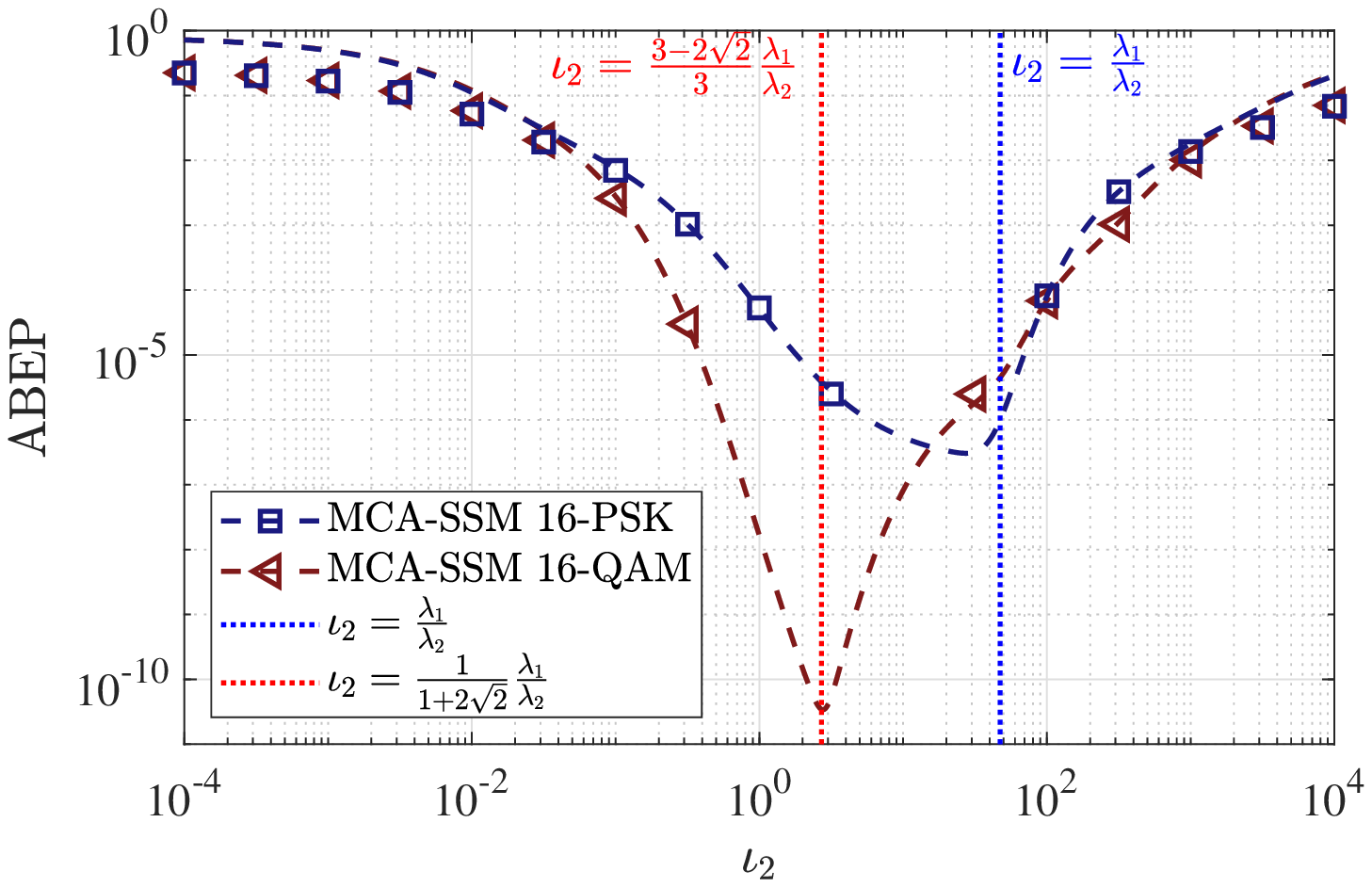}}
\caption{Numerical validation of ABEP.
 Solid lines and dashed lines show the UUB computed by (\ref{eq14}). Markers show simulation results.
}\label{ABEPvalidation}
\end{figure}

ABEPs of the proposed MCA-SSM system with $\iota_{\mathrm{opt},2}$, $\iota_{2}=1$ and $\iota_{2}=100$ are illustrated in Fig. \ref{ABEPvalidation}(a) and Fig. \ref{ABEPvalidation}(b).
 It validates that
 $\iota_{\mathrm{opt},2}
=
\frac{\lambda_1}{\lambda_2}$ and
 $\iota_{\mathrm{opt},2}
=
\frac{(\sqrt{2}-1)^2}{3}
\frac{\lambda_1}{\lambda_2}$, obtained via Theorem \ref{TheoremT1}, achieve the lowest ABEP for Example 2 and Example 3, respectively. Overall, ABEPs for both Example 2 and Example 3 against $\iota_2$ are illustrated in Fig. \ref{ABEPvalidation}(c) to offer a clear validation of Theorem \ref{TheoremT1}. It can be observed that the ABEP is minimized via the proposed approach.

Case studies in this section provide proof-of-principle evaluation for the proposed MCA-SSM system.
Beyond these case studies, there is a need to evaluate and compare the proposed MCA-SSM systems before deployment in a more practical environment.

\section{Numerical analysis in a typical indoor environment}

Evaluation and comparison of the proposed MCA-SSM system considering environmental factors should be taken into account carefully before network deployment. Especially, evaluation in an indoor environment is crucial as most wireless traffics take place indoors \cite{indoor1,indoor2,indoor3,bm1,bm2,bm3}.
In Section V, we have investigated the performance of MCA-SSM with channel parameters listed in TABLE \ref{table_exampparam}. However, channel parameters may vary with different locations of Tx and Rx in a typical building environment.
In this section, we apply the deterministic channel model as a powerful tool to evaluate the performance of the proposed MCA-SSM systems in a typical indoor environment \cite{Zhihua}.

\subsection{Channel prediction}

\begin{figure}[h]
\centering
\subfigure[Simulation environment]
{\includegraphics[scale=0.6]{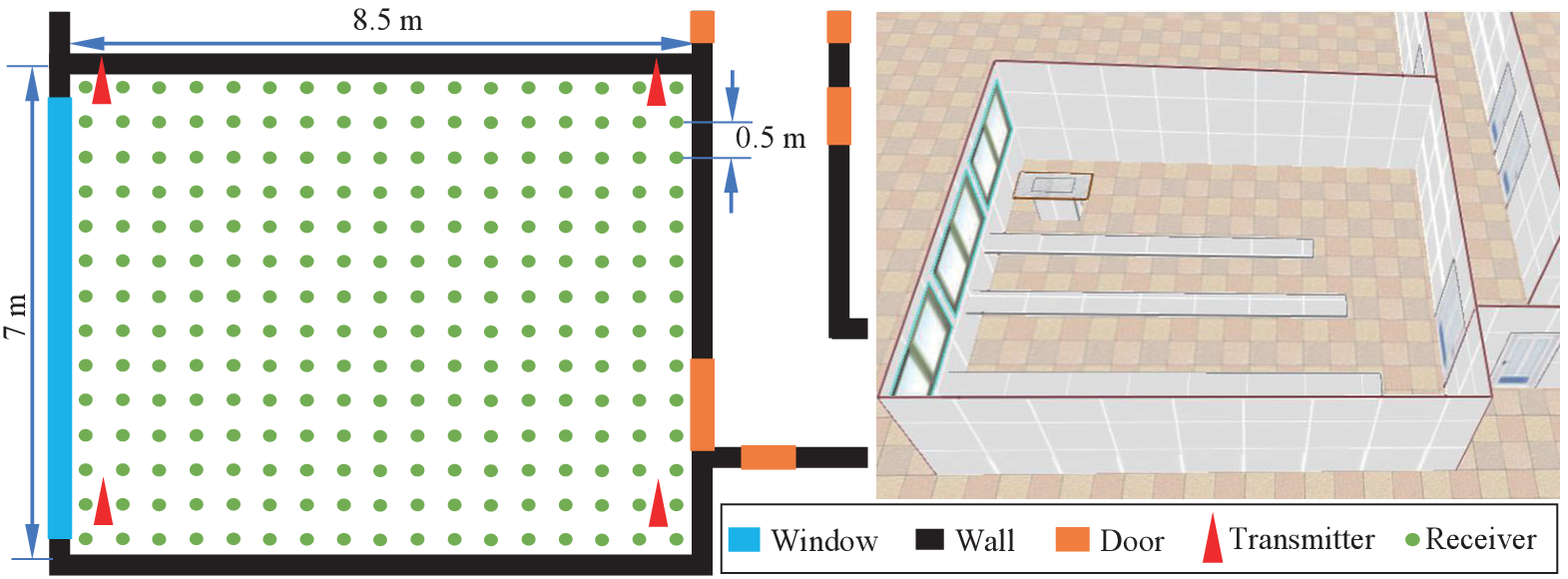}}
\subfigure[A channel prediction instance.]
{\includegraphics[scale=0.6]{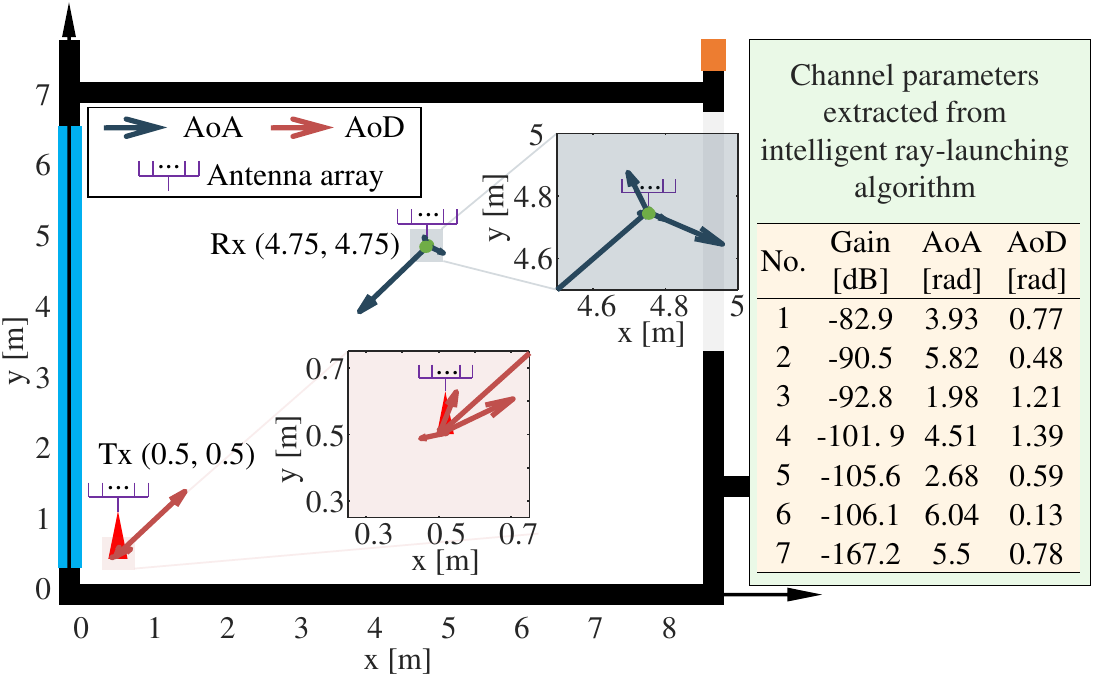}}
\caption{IRLA-based channel prediction in a typical indoor environment.} \label{channelprediction}
\end{figure}

The indoor environment under consideration is illustrated in Fig. \ref{channelprediction}(a). The IRLA, which has been validated via practical measurement \cite{RadioScience,RadioScience2}, is employed to predict indoor propagation channels.

In the prediction, the resolution is set as 0.5 m to achieve a good tradeoff between computational complexity and prediction accuracy. There are four Tx antenna arrays, located at four corners of the room, respectively.
The IRLA slices the propagation environment into small cube units with a dimension of 0.5 m $\times$ 0.5 m $\times$ 0.5 m, and predicts channels between the Tx and centers of all cubes. The height of both Tx and Rx is set as 1 m above the floor. The Rx array visits all cube centers with a height of 1 m and computes the ergodic performance of SSM accordingly.

With 4 Tx positions and 238 Rx positions, there are $N_\mathrm{l}=952$ links in total. The index of a link is denoted by $n_\mathrm{l}\in \{1,2,...,952\}$. 
In each channel prediction, we obtain channel parameters including the number of MP components $N_\mathrm{ts}$, 
 the gain of each MP component ${\bm{\beta}}$, 
AoD of each MP component $\mathbf{\theta}_\mathrm{t}$,
and
AoA of each MP component $\mathbf{\theta}_\mathrm{r}$. 
In the example shown in Fig. \ref{channelprediction}(b), Tx and Rx are located at (0.5,0.5) and (4.75,4.75), respectively. Seven MP components are present in this link. For all the other links, the IRLA is applied to extract channel parameters with the same data format.

\subsection{Numerical results}
The UUB on the ABEP of the proposed MCA-SSM system is systematically analyzed and compared in the indoor environment illustrated in Fig. \ref{channelprediction}.
 In all figures in this section, the solid lines are closed-form ABEP upper bounds computed by (\ref{eq14}), and markers are simulation results. Parameters of both the conventional SSM and the proposed MCA-SSM in this section are provided in TABLE \ref{table_param} unless otherwise specified.
 
 \begin{table}[h]
\centering
\caption{Parameters in numerical results.}
\label{table_param}
\begin{tabular}{l l l}
\hline 
\bf{Parameter} & \bf{Description} & \bf{Value} \\ 
\hline
$N_\mathrm{r}$&Number of Rx antennas&16\\
$N_\mathrm{t}$&Number of Tx antennas&16\\
$d_\mathrm{t}$&Spacing of Tx antenna elements&0.5$\lambda$\\
$d_\mathrm{r}$&Spacing of Rx antenna elements&0.5$\lambda$\\
$N_\mathrm{s}$&Number of candidate MP components&4\\
$L$&Number of candidate combinations of MP components&4\\
$M$&Modulation order&16\\
\hline
\end{tabular}
\end{table}

\begin{figure}[h]
\centering
\includegraphics[scale=0.5]{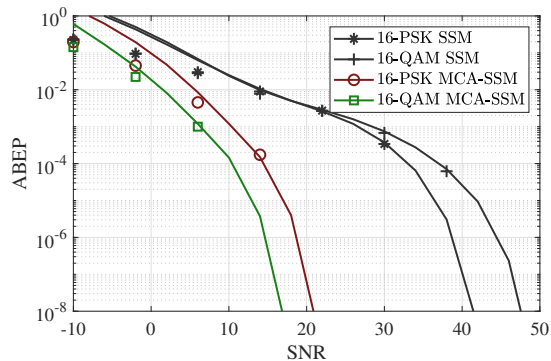}
\caption{Comparison between the proposed MCA-SSM with conventional SSM.
 Solid lines show the UUB computed by (\ref{eq14}). Markers show simulation results.
}\label{f1}
\end{figure}

The ABEP of the proposed MCA-SSM is compared with that of the conventional SSM in Fig. \ref{f1}. It is shown that the proposed MCA-SSM system outperforms the conventional SSM system significantly in terms of ABEP. 
To achieve an ABEP of $10^{-6}$, the proposed MCA-SSM with 16-PSK requires nearly 20 dB
less transmit power than the conventional SSM, while the MCA-SSM with 16-QAM requires nearly 30 dB less.
It reveals that conventional SSM systems suffer from both low cluster gain and
inter-MP interference significantly. On one hand, assumptions in \cite[Eq. (2)]{Spatial} and \cite[Eq. (4)]{Generalized} are not applicable for a MIMO system with relatively small-size antenna arrays. On the other hand, the assumption of ${\bm\beta}\sim\mathcal{CN}(0,\mathbf{I})$, applied in \cite{Spatial,Generalized,Polarized,Adaptive,Diversity}, overestimates the performance of conventional SSM systems. Fortunately, the proposed MCA-SSM systems overcome both problems effectively.

\begin{figure}[h]
\centering
\subfigure[PSK]{\includegraphics[scale=0.5]{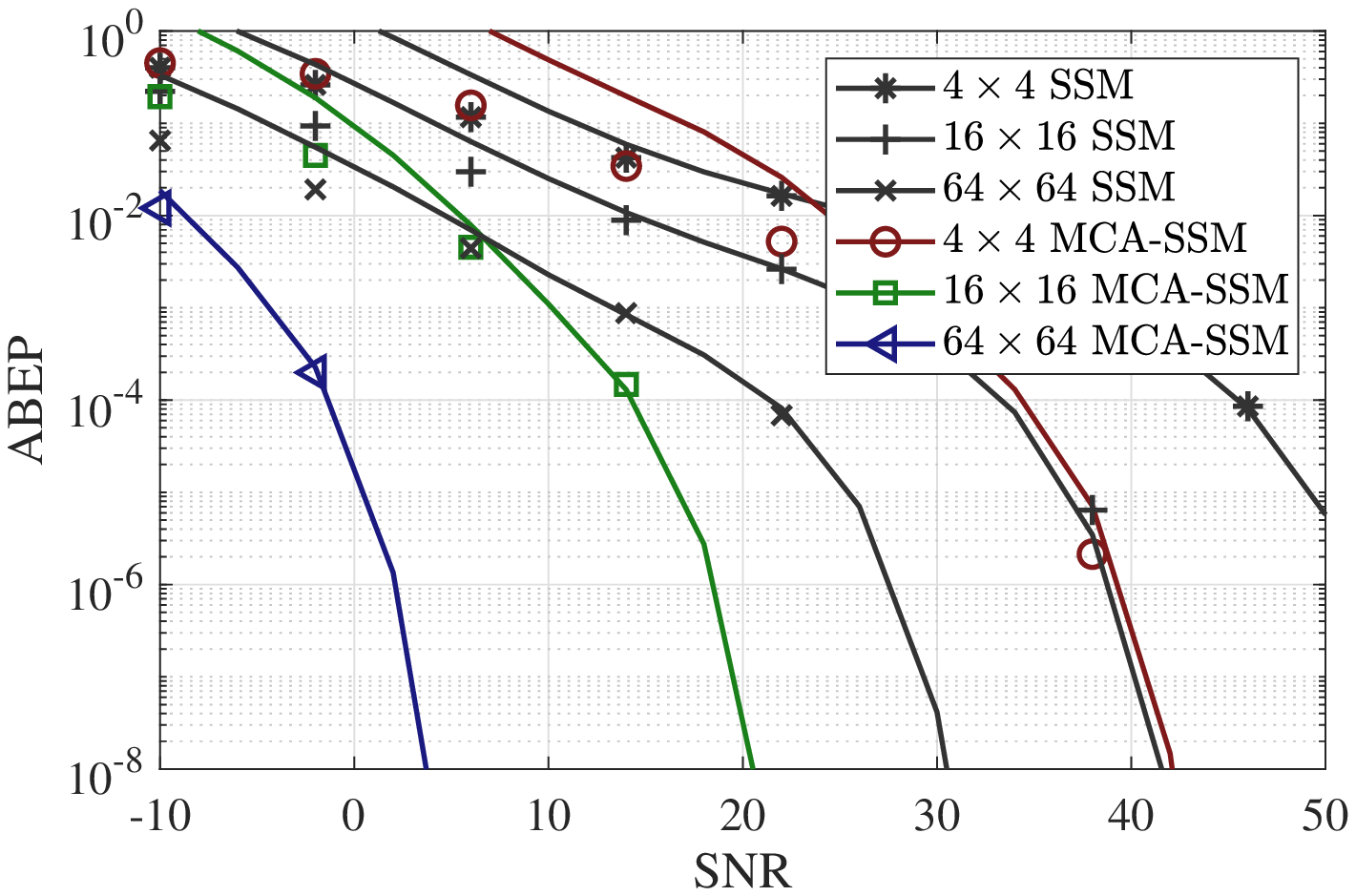}}
\subfigure[QAM]{\includegraphics[scale=0.5]{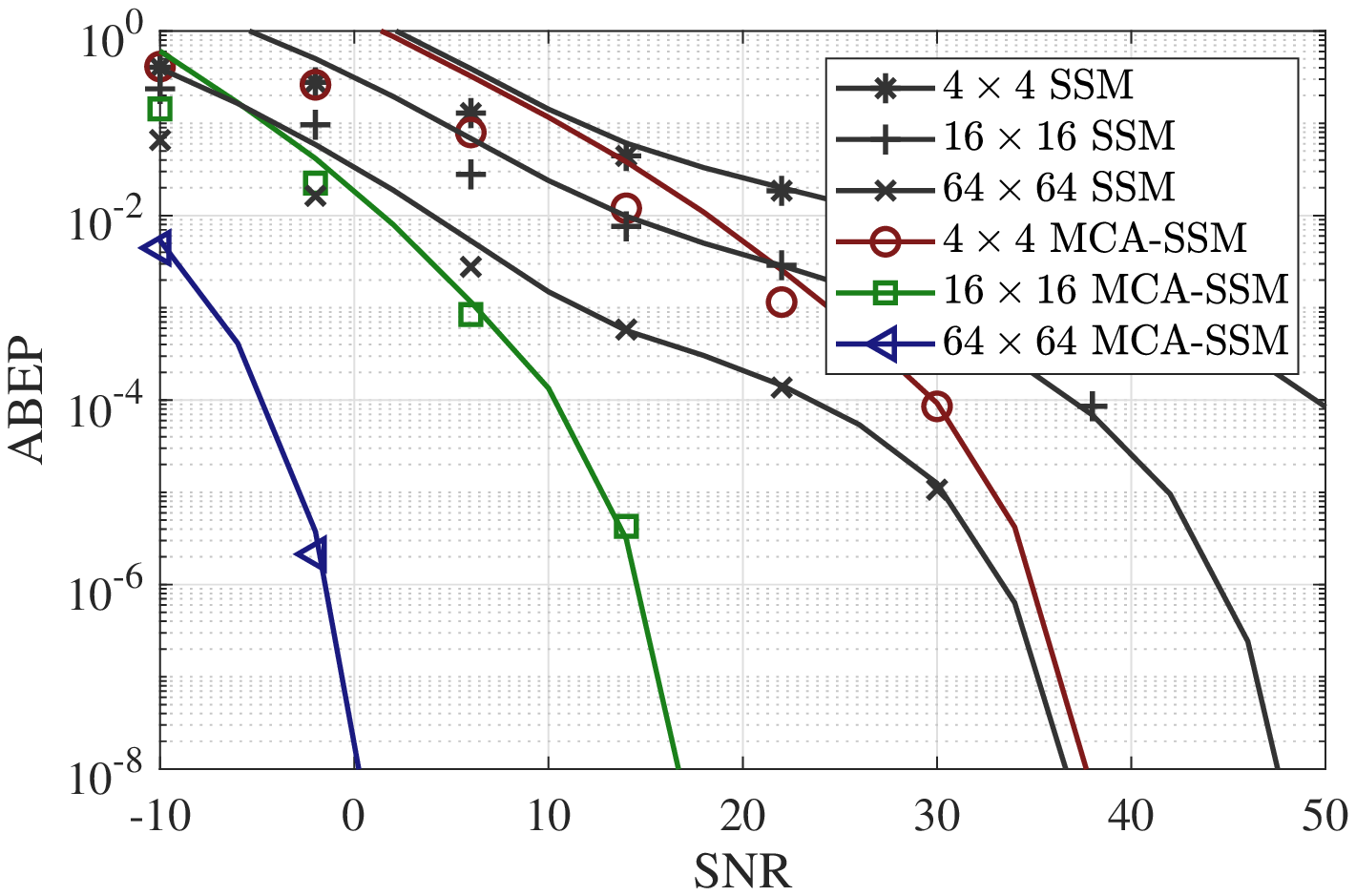}}
\caption{Comparison of MCA-SSM with different scales of antenna arrays, with  solid lines showing the UUB computed by (\ref{eq14}), and markers showing simulation results.
}\label{f2}
\end{figure}

Then, the impact of antenna array size on the ABEP is shown in Fig. \ref{f2}. It can be seen that the ABEP can be effectively reduced by increasing the scale of the antenna array for both  systems. Nevertheless, more gain can be achieved via adding antenna elements for the proposed MCA-SSM system in comparison with the conventional SSM. In Fig. \ref{f2}, the gain of the proposed MCA-SSM and the conventional SSM via scaling up antenna array
from $4\times4$ to $16\times 16$
are, respectively, nearly 20 dB and
10 dB
 at $\mathrm{ABEP}=10^{-6}$.
This result reveals that the bit error performance of conventional SSM suffers from MP components with low gains, even though they are well resolved via large-scale antenna arrays at both the Tx and the Rx.

\begin{figure}[h]
\centering
\includegraphics[scale=0.5]{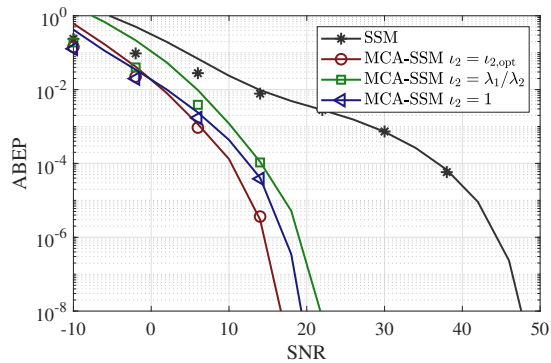}
\caption{Performance of MCA-SSM with optimal $\iota_{2,\mathrm{opt}}$, suboptimal $\iota_2=\lambda_1/\lambda_2$, and suboptimal $\iota_2=1$.
 Solid lines show the UUB computed by (\ref{eq14}). Markers show simulation results.
 }\label{f3}
\end{figure}

Inspired by Fig. \ref{ABEPvalidation}, it seems that the MCA mechanism without the optimisation of $\bm\iota$, where the MCA matrix $\mathbf{W}$ is easy to compute, could still achieve a relatively good bit error performance. To test this, Fig. \ref{f3} illustrates ABEPs of the MCA-SSM with optimal $\iota_{2,\mathrm{opt}}$, $\iota_2=\lambda_1/\lambda_2$, and $\iota_2=1$, respectively. In Fig. \ref{f3}, the 16-QAM signal constellation is applied. Surprisingly, the proposed MCA-SSM system with $\iota_2=\lambda_1/\lambda_2$ and $\iota_2=1$ still significantly outperforms the conventional SSM system. 
It is shown that 
SNR losses due to suboptimal $\iota_2=\lambda_1/\lambda_2$ and $\iota_2=1$ are, respectively, only 2.5 dB and 4.5 dB at $\mathrm{ABEP}=10^{-6}$. Therefore, we can sacrifice the optimality of $\iota_2$ to reduce the implementation complexity in the designed MCA-SSM
 in case the computation resources are rather limited.

\section{Conclusions}
This paper proposes a new SSM system with a novel MCA mechanism introduced
 and the involved MCA matrix is analytically optimized to minimize the ABEP.
The proposed approach is validated via both analytical UUB on the ABEP and Monte-Carlo simulations.
The ABEP of the proposed MCA-SSM system is also systematically evaluated based on indoor channel predictions using IRLA. 
All the results show that the proposed MCA-SSM system outperforms the SSM system significantly, and can be considered as a promising candidate for 5G/B5G modulation systems with massive MIMO antenna arrays at both the transmitter and the receiver.

\appendices

\section{Proof of Theorem \ref{TheoremT1}}
\label{ProveT1}

The objective of solving (\ref{optimisation3}) is to find the optimal ${\bm \xi}$ that maximizes $J_{l_1,m_1,l_2,m_2}$, which is denoted by 
$J_{\mathrm{min}}\triangleq\min\limits_
{
\begin{smallmatrix}
(l_1,m_1)\neq (l_2,m_2)\\
1\leq l_1\leq L, l_1\in \mathbb{N}^+\\
1\leq m_1\leq M, m_1\in \mathbb{N}^+\\
1\leq l_2\leq L, l_2\in \mathbb{N}^+\\
1\leq m_2\leq M, m_2\in \mathbb{N}^+
\end{smallmatrix}}
J_{l_1,m_1,l_2,m_2}$ in this appendix.

Consider
\begin{eqnarray}
{\bm \xi}^2={\bm \xi}^2(1){\bm\iota},
\end{eqnarray}
where
\begin{eqnarray}
{\bm\iota}=[1,{\iota}_2,...,{\iota}_{N_{\mathrm{s,a}}}],
\end{eqnarray}
and
\begin{eqnarray}
{\iota}_\kappa\triangleq\frac{{\bm \xi}^2(\kappa)}{{\bm \xi}^2(1)}.
\end{eqnarray}
Then, as long as 
\begin{eqnarray}
{\bm \xi}^2(1)=\frac{1}{\sum\limits_{\kappa=1}^{N_{\mathrm{s,a}}}{\iota}_\kappa},
\end{eqnarray}
we can directly calculate ${\bm \xi}^2$ with its associate ${\bm\iota}$ using (\ref{eq38}). 
As a result, the ED for given ${\bm\iota}$ can be expressed as
\begin{eqnarray}
J_{l_1,m_1,l_2,m_2}=\frac{\sum\limits_{\kappa=1}^{N_\mathrm{s,a}}{\iota}_\kappa
{\bm\epsilon}_{l_1,m_1,l_2,m_2}(\kappa)
}{\sum\limits_{\kappa=1}^{N_{\mathrm{s,a}}}{\iota}_\kappa},
\end{eqnarray}
where
\begin{eqnarray}
{\bm\epsilon}_{l_1,m_1,l_2,m_2}\triangleq
{\bm\lambda}
\odot
|{\bm \upsilon}_{l_1}
s_{m_1}
-
{\bm \upsilon}_{l_2}
s_{m_2}|^2.
\end{eqnarray}
Therefore, (\ref{optimisation3}) can be reformulated as
\begin{eqnarray}
\label{optimisation4}
\begin{aligned}
\mathrm{maximize} &\ \min\limits_
{
\begin{smallmatrix}
(l,m)\neq (\hat{l},\hat{m})\\
1\leq l\leq L, l\in \mathbb{N}^+\\
1\leq m\leq M, m\in \mathbb{N}^+\\
1\leq \hat{l}\leq L, \hat{l}\in \mathbb{N}^+\\
1\leq \hat{m}\leq M, \hat{m}\in \mathbb{N}^+
\end{smallmatrix}
}
\frac{\sum\limits_{\kappa=1}^{N_\mathrm{s,a}}{\iota}_\kappa
{\bm\epsilon}_{l_1,m_1,l_2,m_2}(\kappa)
}{\sum\limits_{\kappa=1}^{N_{\mathrm{s,a}}}{\iota}_\kappa},
\\
\mathrm{s.t.} &\ {\bm\iota}_{2:N_{\mathrm{s,a}}}\in (\mathbb{R}^+)^{N_{\mathrm{s,a}}-1}.
\end{aligned}
\end{eqnarray}

Note that the value of $J_{l_1,m_1,l_2,m_2}$ is determined by both 
${\bm\epsilon}_{l_1,m_1,l_2,m_2}$ and $\bm\iota$.
There are as many as $(LM)^2-1$ sets of ${\bm\epsilon}_{l_1,m_1,l_2,m_2}$. The following two challenging questions need to be addressed to solve (\ref{optimisation4}).
\begin{itemize}
\item 
\texttt{Challenge 1:}
Which ${\bm\epsilon}_{l_1,m_1,l_2,m_2}$ can possibly form a candidate solution of 
$J_{\mathrm{min}}$ regardless of $\bm\iota$? 
For two values of  ${\bm\epsilon}_{l_1,m_1,l_2,m_2}$, denoted by 
	${\bm\epsilon}_{{l_{1,_\mathrm{A}},m_{1,\mathrm{A}},l_{2,_\mathrm{A}},m_{2,\mathrm{A}}}}$ and 
	${\bm\epsilon}_{{l_{1,_\mathrm{B}},m_{1,\mathrm{B}},l_{2,_\mathrm{B}},m_{2,\mathrm{B}}}}$,
 if 
$J_{{l_{1,_\mathrm{A}},m_{1,\mathrm{A}},l_{2,_\mathrm{A}},m_{2,\mathrm{A}}}}
>J_{{l_{1,_\mathrm{B}},m_{1,\mathrm{B}},l_{2,_\mathrm{B}},m_{2,\mathrm{B}}}}
, \forall {\bm\iota}$, then
$J_{{l_{1,\mathrm{A}},m_{1,\mathrm{A}},l_{2,\mathrm{A}},m_{2,\mathrm{A}}}}$ is not a possible candidate solution. 
\item 
\texttt{Challenge 2:}
How to find the optimal solution of ${\bm\iota}_{\mathrm{opt}}$ that maximizes $J_{\mathrm{min}}$, provided candidate solutions of ${\bm\epsilon}_{l_1,m_1,l_2,m_2}$ calculated by addressing \texttt{Challenge 1}? 
\end{itemize}

\texttt{Challenge 1} is addressed as follows. Note that  $J_{l_1,m_1,l_2,m_2}$ has two types for $l_1=l_2$ and $l_1\neq l_2$, respectively \cite{SMGC}. Then, we have 
\begin{eqnarray}
\label{eq71}
J_{\mathrm{min}}
=\min\{J_{\mathrm{min,Mod}},J_{\mathrm{min,MC}}\},
\end{eqnarray}
where
\begin{eqnarray}
\label{J1def}
J_{\mathrm{min,Mod}}&\triangleq&
\min\limits_
{
\begin{smallmatrix}
m_1\neq m_2\\
1\leq l_1\leq L, l_1\in \mathbb{N}^+\\
1\leq m_1\leq M, m_1\in \mathbb{N}^+\\
1\leq m_2\leq M, m_2\in \mathbb{N}^+
\end{smallmatrix}}
J_{l_1,m_1,l_1,m_2},
\\
\label{J2def}
J_{\mathrm{min,MC}}&\triangleq&
\min\limits_
{
\begin{smallmatrix}
1\leq l_1\leq L, l_1\in \mathbb{N}^+\\
1\leq m_1\leq M, m_1\in \mathbb{N}^+\\
1\leq l_2\leq L, l_2\in \mathbb{N}^+, l_2\neq l_1\\
1\leq m_2\leq M, m_2\in \mathbb{N}^+
\end{smallmatrix}}
J_{l_1,m_1,l_2,m_2}.
\end{eqnarray}

Then the statement ``${\bm\epsilon}_{l_1,m_1,l_2,m_2}$ forms a candidate solution of either 
$J_{\mathrm{min,Mod}}$
or $J_{\mathrm{min,MC}}$.'' is a necessary condition for the statement ``${\bm\epsilon}_{l_1,m_1,l_2,m_2}$ forms a candidate solution of $J_{\mathrm{min}}$.''.
Therefore, candidate solutions of $J_{\mathrm{min,MC}}$ is the union of candidate solutions to both $J_{\mathrm{min,Mod}}$
and $J_{\mathrm{min,MC}}$. 

Candidate solutions of $J_{\mathrm{min,Mod}}$
and $J_{\mathrm{min,MC}}$ with given $\bm\iota$ are derived in Lemma \ref{lemma2} and Lemma \ref{lemma3}, respectively.

\begin{lemma}\label{lemma2} \textbf{Candidate solutions of $J_{\mathrm{min,mod}}$.}
\begin{eqnarray}
\label{ep1}
{\bm\epsilon}_{l_1,m_1,l_1,m_2}\in
\left\{
({\bm\lambda}\odot|{\bm \upsilon}_{l_1}
|^2)\times
\left\{
\begin{array}{ll}
4\sin^2\left(\frac{\pi}{M}\right),& \mathrm{PSK},\\
\frac{6}{M-1},& \mathrm{square\ QAM},
\\
\frac{24}{5M-4},& \mathrm{rectangular\ QAM}.
\end{array}
\right.
\right\}.
\end{eqnarray}
forms candidate solutions of $J_{\mathrm{min,mod}}$,
and
\begin{eqnarray}
\label{eq56}
J_{\mathrm{min,mod}}=\min\limits_
{
\begin{smallmatrix}
1\leq l_1\leq L, l_1\in \mathbb{N}^+\\
\end{smallmatrix}}
\left\{
\sum_{\kappa=1}^{N_\mathrm{s}}{\bm \xi}^2(\kappa)\lambda_\kappa
|
{\bm \upsilon}_{l_1}(\kappa)
|^2\times
\left\{
\begin{array}{ll}
4\sin^2\left(\frac{\pi}{M}\right),& \mathrm{PSK},\\
\frac{6}{M-1},& \mathrm{square\ QAM},
\\
\frac{24}{5M-4},& \mathrm{rectangular\ QAM}.
\end{array}
\right.\right\}
\end{eqnarray}
\end{lemma}
\begin{IEEEproof}
See Appendix \ref{prooflemma2}.

\end{IEEEproof}

\begin{lemma}\label{lemma3} \textbf{Candidate solutions of $J_{\mathrm{min,MC}}$.}
\begin{eqnarray}
\label{ep2}
{\bm\epsilon}_{l_1,m_1,l_2,m_2}
\in
\left\{
\begin{array}{ll}
\left\{{\bm\lambda}\odot
|{\bm \upsilon}_{{l_1}}\pm{\bm \upsilon}_{{l_2}}|^2\right\},
& \mathrm{PSK},\\
\left\{|{\bm \upsilon}_{{l_1}}
s_{{m_1}}
\pm
{\bm \upsilon}_{{l_2}}
s_{m_2}|^2,
(s_{m_1},s_{m_2})\in\mathcal{S}_{\mathrm{m}}\right\}, & \mathrm{QAM},
\end{array}
\right.
\end{eqnarray}
forms candidate solutions of $J_{\mathrm{min,MC}}$, where the sets
$\mathcal{S}_{\mathrm{m}}$ for square QAM and rectangular QAM signal constellations 
are calculated by (\ref{smpsk}) and (\ref{smqam}), respectively.

As such
\begin{eqnarray}
\label{eq60}
J_{\mathrm{min,MC}}
=
\min\limits_
{
\begin{smallmatrix}
1\leq l_1\leq L, l_1\in \mathbb{N}^+\\
1\leq l_2\leq L, l_2\in \mathbb{N}^+, l_2\neq l_1
\end{smallmatrix}}
\left\{
\begin{array}{ll}
\min\left\{
\sum\limits_{\kappa=1}^{N_\mathrm{s,a}}{\bm \xi}^2(\kappa)\lambda_\kappa
({\bm \upsilon}_{{l_1}}(\kappa)\pm{\bm \upsilon}_{{l_2}}(\kappa))^2
\right\}
,& \mathrm{PSK},\\
\min\limits_
{
\begin{smallmatrix}
(s_{m_1},s_{m_2})\in\mathcal{S}_{\mathrm{m}}
\end{smallmatrix}}
\left\{
\sum\limits_{\kappa=1}^{N_\mathrm{s,a}}{\bm \xi}^2(\kappa)\lambda_\kappa
|
{\bm \upsilon}_{{l_1}}(\kappa)
s_{{m_1}}
\pm
{\bm \upsilon}_{{l_2}}(\kappa)
s_{m_2}|^2\right\},& \mathrm{QAM}.
\end{array}
\right.
\end{eqnarray}

\end{lemma}
\begin{IEEEproof}
See Appendix \ref{prooflemma3}
\end{IEEEproof}

We can see the optimal solution of (\ref{optimisation4})
must be formed by an $\bm\epsilon$ in either set (\ref{ep1}) or set (\ref{ep2}), and thus the candidate $\bm\epsilon$ is in the union of both sets, which is denoted by $\mathcal{D}_0$ and summarized in 
(\ref{D1}) and (\ref{D2}).

Moreover, for a given ${\bm\epsilon}\in\mathcal{D}_0$, if ${\bm\epsilon}(\kappa)\geq {\bm\epsilon}'(\kappa),\forall\kappa, \exists{\bm\epsilon}'\in\mathcal{D}_0,{\bm\epsilon}'\neq{\bm\epsilon}$, ${\bm\epsilon}$ always forms a greater $J_{l_1,m_1,l_2,m_2}$ than ${\bm\epsilon}'$, and can not be the solution of (\ref{eq71}). Therefore, it needs to be excluded from the candidate set of ${\bm\epsilon}$, and thus we have (\ref{eq34}).

\texttt{Challenge 2} is addressed via the following Lemma \ref{lemma1}. 

\begin{lemma}\label{lemma1}\textbf{Monotonicity of $J_{l_1,m_1,l_2,m_2}$.}
$J_{l_1,m_1,l_2,m_2}$ is monotonous to ${\iota}_{\kappa_0},\forall1<\kappa_0\leq N_{\mathrm{s,a}}$, for a given $(l_1,m_1,l_2,m_2)$.
\end{lemma}
\begin{IEEEproof}
See Appendix \ref{prooflemma1}.
\end{IEEEproof}

Due to the monotonicity, the optimal solution of ${\bm\iota}$, which contains $N_{\mathrm{s,a}}-1$ unknown numbers, surely be one of intersections of $N_{\mathrm{s,a}}-1$ sets of $J_{l_1,m_1,l_2,m_2}$ out of all candidates in (\ref{eq56}) and (\ref{eq60}). Then, the intersection of $J_{l_1,m_1,l_2,m_2}={\bm\epsilon}_{l_1,m_1,l_2,m_2}{\bm\iota}^\mathrm{T}$ can be computed by (\ref{eq33}). 

After addressing \texttt{Challenge 1} and \texttt{Challenge 2}, Theorem \ref{TheoremT1} is proved.

\section{Proof of Lemma \ref{lemma2}}
\label{prooflemma2}

According to (\ref{J1def}), we have
\begin{eqnarray}
\label{eq58}
\begin{aligned}
J_{\mathrm{min,mod}}&=&&
\min\limits_
{
\begin{smallmatrix}
m_1\neq m_2\\
1\leq l_1\leq L, l_1\in \mathbb{N}^+\\
1\leq m_1\leq M, m_1\in \mathbb{N}^+\\
1\leq m_2\leq M, m_2\in \mathbb{N}^+
\end{smallmatrix}}
\left\{
\sum_{\kappa=1}^{N_\mathrm{s}}{\bm \xi}^2(\kappa)\lambda_\kappa
|
{\bm \upsilon}_{l_1}(\kappa)
s_{m_1}
-
{\bm \upsilon}_{l_1}(\kappa)
s_{m_2}|^2\right\}
\\
&=&&\min\limits_
{
\begin{smallmatrix}
m_1\neq m_2\\
1\leq m_1\leq M, m_1\in \mathbb{N}^+\\
1\leq m_2\leq M, m_2\in \mathbb{N}^+
\end{smallmatrix}}
\{|
s_{m_1}
-
s_{m_2}|^2\}
\min\limits_
{
\begin{smallmatrix}
1\leq l_1\leq L, l_1\in \mathbb{N}^+\\
\end{smallmatrix}}
\left\{
\sum_{\kappa=1}^{N_\mathrm{s}}{\bm \xi}^2(\kappa)\lambda_\kappa
|
{\bm \upsilon}_{l_1}(\kappa)
|^2\right\}.
\end{aligned}
\end{eqnarray}

For typical APM modulations, i.e., PSK and QAM, we have
\begin{eqnarray}
\label{eq77}
\min\limits_
{
\begin{smallmatrix}
m\neq \hat{m}\\
1\leq m\leq M, m\in \mathbb{N}^+\\
1\leq \hat{m}\leq M, \hat{m}\in \mathbb{N}^+
\end{smallmatrix}
}
\{|
s_{\hat{m}}
-
s_{m}|\}=
\left\{
\begin{array}{ll}
2\sin\left(\frac{\pi}{M}\right)
,&\mathrm{PSK},\\
\sqrt{\frac{6}{M-1}}
,&\mathrm{square\ QAM},\\
\sqrt{\frac{24}{5M-4}}
,&\mathrm{rectangular\ QAM}.
\end{array}
\right.
\end{eqnarray}
By substituting (\ref{eq77})
into (\ref{eq58}), we obtain (\ref{eq56}).

\section{Proof of Lemma \ref{lemma3}}
\label{prooflemma3}
According to the definition of $J_{\mathrm{min,MC}}$ in (\ref{J2def}),
\begin{eqnarray}
\label{eq81}
J_{\mathrm{min,MC}}=
\min\limits_
{
\begin{smallmatrix}
1\leq l_1\leq L, l_1\in \mathbb{N}^+\\
1\leq l_2\leq L, l_2\in \mathbb{N}^+, l_2\neq l_1
\end{smallmatrix}}
\min\limits_
{
\begin{smallmatrix}
1\leq m_1\leq M, m_1\in \mathbb{N}^+\\
1\leq m_2\leq M, m_2\in \mathbb{N}^+
\end{smallmatrix}}\left\{
\sum_{\kappa=1}^{N_\mathrm{s}}{\bm \xi}^2(\kappa)\lambda_\kappa
|
{\bm \upsilon}_{{l_1}}(\kappa)
s_{{m_1}}
-
{\bm \upsilon}_{{l_2}}(\kappa)
s_{m_2}|^2\right\}.
\end{eqnarray}

Note that
\begin{eqnarray}
\label{eq64}
\begin{aligned}
\sum_{\kappa=1}^{N_\mathrm{s,a}}{\bm \xi}^2(\kappa)\lambda_\kappa
|
{\bm \upsilon}_{{l_1}}(\kappa)
s_{{m_1}}
-
{\bm \upsilon}_{{l_2}}(\kappa)
s_{m_2}|^2
&=&&
|s_{{m_1}}|^2\sum_{\kappa=1}^{N_\mathrm{s,a}}{\bm \xi}^2(\kappa)\lambda_\kappa
{\bm \upsilon}_{{l_1}}^2(\kappa)
\\
&&&+
|s_{{m_2}}|^2\sum_{\kappa=1}^{N_\mathrm{s,a}}{\bm \xi}^2(\kappa)\lambda_\kappa
{\bm \upsilon}_{{l_2}}^2(\kappa)
\\
&&&-2\mathcal{R}[s_{m_1}s_{m_2}^*]
\sum_{\kappa=1}^{N_\mathrm{s,a}}{\bm \xi}^2(\kappa)\lambda_\kappa
{\bm \upsilon}_{{l_1}}(\kappa)
{\bm \upsilon}_{{l_2}}(\kappa).
\end{aligned}
\end{eqnarray}

Here, we work on (\ref{eq64}) for PSK and QAM separately.

For PSK, $|s_{{m_1}}|^2=|s_{{m_2}}|^2=1$, and $\mathcal{R}[s_{m_1}s_{m_2}^*]=\cos\left(\frac{2\pi(m_1-m_2)}{M}\right)$. Thus, we have 
\begin{eqnarray}
\label{eq64PSK}
\begin{aligned}
\sum_{\kappa=1}^{N_\mathrm{s,a}}{\bm \xi}^2(\kappa)\lambda_\kappa
|
{\bm \upsilon}_{{l_1}}(\kappa)
s_{{m_1}}
-
{\bm \upsilon}_{{l_2}}(\kappa)
s_{m_2}|^2
&=&&
\sum_{\kappa=1}^{N_\mathrm{s,a}}{\bm \xi}^2(\kappa)\lambda_\kappa
{\bm \upsilon}_{{l_1}}^2(\kappa)
+
\sum_{\kappa=1}^{N_\mathrm{s,a}}{\bm \xi}^2(\kappa)\lambda_\kappa
{\bm \upsilon}_{{l_2}}^2(\kappa)
\\
&&&-2\cos\left(\frac{2\pi(m_1-m_2)}{M}\right)
\sum_{\kappa=1}^{N_\mathrm{s,a}}{\bm \xi}^2(\kappa)\lambda_\kappa
{\bm \upsilon}_{{l_1}}(\kappa)
{\bm \upsilon}_{{l_2}}(\kappa).
\end{aligned}
\end{eqnarray}

From (\ref{eq64PSK}), we see that $\sum_{\kappa=1}^{N_\mathrm{s,a}}{\bm \xi}^2(\kappa)\lambda_\kappa
|
{\bm \upsilon}_{{l_1}}(\kappa)
s_{{m_1}}
-
{\bm \upsilon}_{{l_2}}(\kappa)
s_{m_2}|^2$ is minimized by 
$\frac{2\pi(m_1-m_2)}{M}=0$ and $\frac{2\pi(m_1-m_2)}{M}=\pi$ while $\sum\limits_{\kappa=1}^{N_\mathrm{s,a}}{\bm \xi}^2(\kappa)\lambda_\kappa
{\bm \upsilon}_{{l_1}}(\kappa)
{\bm \upsilon}_{{l_2}}(\kappa)>0$ and $\sum\limits_{\kappa=1}^{N_\mathrm{s,a}}{\bm \xi}^2(\kappa)\lambda_\kappa
{\bm \upsilon}_{{l_1}}(\kappa)
{\bm \upsilon}_{{l_2}}(\kappa)\leq0$, respectively.

Then, we conclude that for the PSK signal constellation,
\begin{eqnarray}
\label{eq68}
\begin{aligned}
&\min\limits_
{
\begin{smallmatrix}
1\leq m_1\leq M, m_1\in \mathbb{N}^+\\
1\leq m_2\leq M, m_2\in \mathbb{N}^+
\end{smallmatrix}}
\left\{
\sum_{\kappa=1}^{N_\mathrm{s}}{\bm \xi}^2(\kappa)\lambda_\kappa
|
{\bm \upsilon}_{{l_1}}(\kappa)
s_{{m_1}}
-
{\bm \upsilon}_{{l_2}}(\kappa)
s_{m_2}|^2
\right\}
\\
&
=
\left\{
\begin{array}{ll}
\sum\limits_{\kappa=1}^{N_\mathrm{s,a}}{\bm \xi}^2(\kappa)\lambda_\kappa
({\bm \upsilon}_{{l_1}}(\kappa)-{\bm \upsilon}_{{l_2}}(\kappa))^2,
& \mathrm{if}
\sum\limits_{\kappa=1}^{N_\mathrm{s,a}}{\bm \xi}^2(\kappa)\lambda_\kappa
{\bm \upsilon}_{{l_1}}(\kappa)
{\bm \upsilon}_{{l_2}}(\kappa)>0,
\\
\sum\limits_{\kappa=1}^{N_\mathrm{s,a}}{\bm \xi}^2(\kappa)\lambda_\kappa
({\bm \upsilon}_{{l_1}}(\kappa)+{\bm \upsilon}_{{l_2}}(\kappa))^2,
& \mathrm{if}
\sum\limits_{\kappa=1}^{N_\mathrm{s,a}}{\bm \xi}^2(\kappa)\lambda_\kappa
{\bm \upsilon}_{{l_1}}(\kappa)
{\bm \upsilon}_{{l_2}}(\kappa)\leq0.
\end{array}
\right.
\end{aligned}
\end{eqnarray}
By substituting (\ref{eq68}) into (\ref{eq81}), we obtain the first row of (\ref{eq60}).

For QAM, we consider the following factor to reduce the search space.
\begin{itemize}
\item In a square $M$-QAM constellation diagram, there are $\sqrt{M}/4+M/8$ valid values of the amplitude of a constellation point. In a rectangular $M$-QAM constellation diagram, there are $\sqrt{M/32}+3M/16$ valid values of the amplitude of a constellation point. They are indexed by $\eta$.
\item For given values of amplitudes of two symbols, (\ref{eq64}) is minimized by minimizing the angle between two straight lines. The first straight line
is formed by $s_{m_1}$ and the origin in the constellation diagram, and the second one is formed by $s_{m_2}$ and the origin. 
\item For a given angle between two straight lines of the two symbols, (\ref{eq64}) is minimized by minimizing the amplitudes of the two symbols.
\end{itemize}

Therefore, we can summarize possible pairs of $s_1$ and $s_2$ that minimize (\ref{eq64}) as
\begin{eqnarray}
\mathcal{S}_{\mathrm{m}}=
\left\{(s_1,s_2)\left|
\begin{smallmatrix}
0<\angle(s_1)\leq \pi/4, \angle(s_1)\leq\angle(s_2)\leq\angle(s_1)+\pi,\\ 
(
(|s_1^*|,|s_2^*|)=(|s_1|,|s_2|)\land
(\sin(\angle(s_2)-\angle(s_1))<\sin(\angle(s_2^*)-\angle(s_1^*)))
)
\\
\lor
(
((|s_1^*|<|s_1|)\lor(|s_2^*|<|s_2|))\land
(\sin(\angle(s_2)-\angle(s_1))=\sin(\angle(s_2^*)-\angle(s_1^*)))
) \\
\not\exists (s_1^*,s_2^*)
\end{smallmatrix}
\right.\right\}.
\end{eqnarray} 

Following some straightforward derivations, we obtain (\ref{smpsk}) and (\ref{smqam}) for the
square QAM and the rectangular QAM signal constellations, respectively.

Therefore,
\begin{eqnarray}
\label{eq70}
\begin{aligned}
\min\limits_
{
\begin{smallmatrix}
1\leq m_1\leq M, m_1\in \mathbb{N}^+\\
1\leq m_2\leq M, m_2\in \mathbb{N}^+
\end{smallmatrix}}
\left\{
\sum_{\kappa=1}^{N_\mathrm{s}}{\bm \xi}^2(\kappa)\lambda_\kappa
|
{\bm \upsilon}_{{l_1}}(\kappa)
s_{{m_1}}
\pm
{\bm \upsilon}_{{l_2}}(\kappa)
s_{m_2}|^2\right\}
\\=
\min\limits_
{
\begin{smallmatrix}
(s_{m_1},s_{m_2})\in\mathcal{S}_{\mathrm{m}}
\end{smallmatrix}}
\left\{
\sum_{\kappa=1}^{N_\mathrm{s}}{\bm \xi}^2(\kappa)\lambda_\kappa
|
{\bm \upsilon}_{{l_1}}(\kappa)
s_{{m_1}}
\pm
{\bm \upsilon}_{{l_2}}(\kappa)
s_{m_2}|^2\right\}.
\end{aligned}
\end{eqnarray}

By substituting (\ref{eq70}) into (\ref{eq81}), we obtain the second row of (\ref{eq60}).

\section{Proof of Lemma \ref{lemma1}}
\label{prooflemma1}
To prove the monotonicity of $J_{l_1,m_1,l_2,m_2;\kappa_0}$ to ${\iota}_{\kappa_0}$, we check $\frac{\partial J}{\partial {\iota}_{\kappa_0}}$.

\begin{eqnarray}
\begin{aligned}
\frac{\partial J}{\partial {\iota}_{\kappa_0}}
&=&&\frac{\partial 
\frac{\sum\limits_{\kappa=1}^{N_\mathrm{s,a}}{\iota}_\kappa\lambda_\kappa
|
{\bm \upsilon}_{l_1}(\kappa)
s_{m_1}
-
{\bm \upsilon}_{l_2}(\kappa)
s_{m_2}|^2}{\sum\limits_{\kappa=1}^{N_{\mathrm{s,a}}}{\iota}_\kappa}}{\partial {\iota}_{\kappa_0}}
\\
&=&&{\bm \xi}^4(1)
\left[
\begin{array}{l}
\lambda_{\kappa_0}
|
{\bm \upsilon}_{l_1}(\kappa_0)
s_{m_1}
-
{\bm \upsilon}_{l_2}(\kappa_0)
s_{m_2}|^2\sum\limits_{\kappa=1}^{N_{\mathrm{s,a}}}{\iota}_\kappa
\\-{\sum\limits_{\kappa=1}^{N_\mathrm{s,a}}{\iota}_\kappa\lambda_\kappa
|
{\bm \upsilon}_{l_1}(\kappa)
s_{m_1}
-
{\bm \upsilon}_{l_2}(\kappa)
s_{m_2}|^2}\\
\end{array}
\right]\\
&=&&
{\bm \xi}^4(1)
\sum_{\kappa\neq\kappa_0=1}^{N_{\mathrm{s,a}}}{\iota}_\kappa
\left(
\begin{array}{l}
\lambda_{\kappa_0}
|
{\bm \upsilon}_{l_1}(\kappa_0)
s_{m_1}
-
{\bm \upsilon}_{l_2}(\kappa_0)
s_{m_2}|^2\\
-\lambda_\kappa
|
{\bm \upsilon}_{l_1}(\kappa)
s_{m_1}
-
{\bm \upsilon}_{l_2}(\kappa)
s_{m_2}|^2
\end{array}
\right).
\end{aligned}
\end{eqnarray}

As 
$
\sum\limits_{\kappa\neq\kappa_0=1}^{N_{\mathrm{s,a}}}{\iota}_\kappa
\left(
\begin{array}{l}
\lambda_{\kappa_0}
|
{\bm \upsilon}_{l_1}(\kappa_0)
s_{m_1}
-
{\bm \upsilon}_{l_2}(\kappa_0)
s_{m_2}|^2\\
-\lambda_\kappa
|
{\bm \upsilon}_{l_1}(\kappa)
s_{m_1}
-
{\bm \upsilon}_{l_2}(\kappa)
s_{m_2}|^2
\end{array}
\right)
$
is irrelevant to ${\iota}_{\kappa_0}$, we have
\begin{eqnarray}
\frac{\partial J}{\partial {\iota}_{\kappa_0}}
\left\{
\begin{array}{l l}
\geq0,&\mathrm{if}
\sum\limits_{\kappa\neq\kappa_0=1}^{N_{\mathrm{s,a}}}{\iota}_\kappa
\left(
\begin{array}{l}
\lambda_{\kappa_0}
|
{\bm \upsilon}_{l_1}(\kappa_0)
s_{m_1}
-
{\bm \upsilon}_{l_2}(\kappa_0)
s_{m_2}|^2\\
-\lambda_\kappa
|
{\bm \upsilon}_{l_1}(\kappa)
s_{m_1}
-
{\bm \upsilon}_{l_2}(\kappa)
s_{m_2}|^2
\end{array}
\right)\geq0,
\\
<0,&\mathrm{if}
\sum\limits_{\kappa\neq\kappa_0=1}^{N_{\mathrm{s,a}}}{\iota}_\kappa
\left(
\begin{array}{l}
\lambda_{\kappa_0}
|
{\bm \upsilon}_{l_1}(\kappa_0)
s_{m_1}
-
{\bm \upsilon}_{l_2}(\kappa_0)
s_{m_2}|^2\\
-\lambda_\kappa
|
{\bm \upsilon}_{l_1}(\kappa)
s_{m_1}
-
{\bm \upsilon}_{l_2}(\kappa)
s_{m_2}|^2
\end{array}
\right)<0.
\end{array}
\right.
\end{eqnarray}
Therefore, $J_{l_1,m_1,l_2,m_2;{\iota}_\kappa}$ is monotonous to ${\iota}_{\kappa_0},\forall1<\kappa_0\leq N_{\mathrm{s,a}}$, for a given $(l_1,m_1,l_2,m_2)$.

\end{document}